\documentclass[12pt]{article}

\pdfoutput=1
\usepackage{jheppub} 
\usepackage{hyperref}
\usepackage{graphicx}
\usepackage{amsmath}
\usepackage{indentfirst}
\usepackage{epsf}
\usepackage{epsfig}
\usepackage[makeroom]{cancel}
\usepackage{caption}
\usepackage{amssymb}

\newcommand{\be}{\begin{equation}}
\newcommand{\ee}{\end{equation}}
\newcommand{\bear}{\begin{eqnarray}}
\newcommand{\ear}{\end{eqnarray}}
\newcommand{\ba}{\begin{array}}
\newcommand{\ea}{\end{array}}

\makeatletter
\def\gsim{\compoundrel>\over\sim}
\def\lsim{\compoundrel<\over\sim}
\def\compoundrel#1\over#2{\mathpalette\compoundreL{{#1}\over{#2}}}
\def\compoundreL#1#2{\compoundREL#1#2}
\def\compoundREL#1#2\over#3{\mathrel
         {\vcenter{\hbox{$\m@th\buildrel{#1#2}\over{#1#3}$}}}}
\makeatother

\preprint{
  \begin{flushright}
    INR-TH-2018-016\\
  \end{flushright}
  }
\title{Numerical study of multiparticle scattering in $\lambda\phi^4$ theory}

\author[a,b]{S.V. Demidov,}
\author[a,b]{B.R. Farkhtdinov}

\affiliation[a]{Institute for Nuclear Research of the Russian Academy of Sciences,\\ 60th October Anniversary prospect 7a, Moscow 117312, Russia}
\affiliation[b]{Moscow Institute of Physics and Technology, \\ Institutsky per. 9, Dolgoprudny 141700, Russia}

\emailAdd{demidov@ms2.inr.ac.ru}
\emailAdd{farkhtdinov@phystech.edu}

\abstract{
  We study numerically classical collisions of waves in
  $\lambda\phi^4$ theory. These processes correspond to multiparticle
  scattering in the semiclassical regime. Parameterizing initial
  and final wavepackets by energy $E$ and particle numbers $N_{i}$, 
  $N_{f}$ we find classically allowed region in the parameter
  space. We describe properties of the scattering solutions at
  the boundary of the classically allowed region. We comment on the
  implications of our results for multiparticle production in the quantum  
  regime. 
}


\begin{document}
\maketitle
\toccontinuoustrue

\section{Introduction}
Multiparticle production in weakly coupled bosonic field theories has
long been of great interest (see
e.g.~\cite{Voloshin:1994yp,Libanov:1997nt} for reviews), which was
initially related to instanton-induced
processes~\cite{Ringwald:1989ee,Espinosa:1989qn}. 
Tree level calculations performed in $(3+1)$-dimensional scalar
$\lambda\phi^4$ theory
show~\cite{Cornwall:1990hh,Goldberg:1990qk,Voloshin:1992mz,Brown:1992ay,Argyres:1992np}  
that multiparticle amplitudes of $1\to N_f$ processes near the threshold 
grow factorially with the number of produced particles $N_f$. This
behaviour persists at one-loop
level~\cite{Voloshin:1992nu} and violates 
unitarity at sufficiently large multiplicities $N_f\sim 1/\lambda$.
It was observed~\cite{Libanov:1994ug}, that corresponding
multiparticle cross section of $1\to N_f$ processes can be
conveniently written in the following functional form
\be
\label{0_1}
\sigma_{1\to N_f} \sim {\rm exp}\left(\frac{1}{\lambda}F(\lambda
N_f, {\cal E})\right) 
\ee
 in the limit $\lambda\to 0$ with $\lambda N_f$ and ${\cal
  E}$ fixed, where ${\cal E}$ is the average kinetic energy of the final
particles. Inspired by this exponential behaviour several
semiclassical techniques for calculation of the multiparticle cross
sections were developed
in~\cite{Khlebnikov:1990ue,Rubakov:1991fb,Son:1995wz,Bezrukov:1995ta,Bezrukov:1999kb}.
They involved singular solutions of classical equations of motion and
allowed to extend the observations obtained with perturbative
calculations to higher energies. Still in spite of those vast
  efforts the most reliable results were
obtained in the regime of small $\lambda N_f$ where corresponding
cross sections turn out to be exponentially suppressed. Unitarity
bounds~\cite{Zakharov:1991rp,Rubakov:1995hq,Libanov:1997nt} indicate
that such suppression of the probability~\eqref{0_1} takes place at
all particle numbers. Interesting insights come from studies of an 
analogue of the scattering amplitudes in $(0+1)$-dimensional theory,
i.e. for the anharmonic oscillator with a quartic
potential~\cite{Bachas:1991fd,Cornwall:1993rh,Jaeckel:2018ipq}, where 
it was found that the perturbative factorial growth is replaced by the
exponential suppression for $\lambda N_f\gsim 1$.

Let us note that recently an interest to the problem of multiparticle
production has been renewed~\cite{Khoze:2014zha, Khoze:2014kka,
  Jaeckel:2014lya, Khoze:2015yba, Degrande:2016oan, Khoze:2017tjt,
  Khoze:2017ifq, Voloshin:2017flq, Khoze:2017lft,
  Khoze:2017uga,Khoze:2018kkz}.
In particular, calculations of
Refs.~\cite{Khoze:2017ifq,Khoze:2018kkz} performed in $\lambda\phi^4$ 
theory with spontaneously broken ${\mathbb Z}_2$ symmetry, exhibited
that, contrary to the common belief, the probability of $2\to N_f$
scattering process near the threshold is not suppressed  at $\lambda
N_f\gg 1$ but exhibits an exponential growth at least near the threshold. 
This conclusion was used to invite the Higgsplosion scenario for
solution of the hierarchy and fine-tuning problems of the Standard
Model. Very recently authors of
Refs.~\cite{Belyaev:2018mtd,Monin:2018cbi} argued that the  
Higgsplosion scenario and in particular the exponential growth of the
multiparticle production probability is not consistent with basic
principles of quantum field theory and at least requires some
modification. Therefore behaviour of the probability~\eqref{0_1} in
$\lambda\phi^4$ field theory (in broken and unbroken phases) at large
$\lambda N_f$ is still an open problem even in the weak coupling
regime which will be assumed in the paper. 

It was long ago understood that the processes $few \to N_f$ for large
$N_f$ can be studied starting from the processes $N_i \to N_f$ where
both initial $N_i$ and final $N_f$  particle numbers are large, i.e. of order
$1/\lambda$. In this case, one can consider classical counterpart of
the quantum scattering processes, i.e. collisions of classical wave
packets. If the ingoing and outgoing waves are in the linear regime at 
$t\to\pm\infty$ corresponding initial and final field configurations
can be associated with coherent states having average energy $E$,
initial $N_i$ and final $N_f$ particle numbers. The probability of
corresponding quantum scattering in the semiclassical regime is not
exponentially suppressed and the whole family of such solutions span
a classically allowed region  in the space of parameters $E, N_i$ and
$N_f$. At a given energy $E_*$ and final particle 
number $N_f$ one can try to minimize the initial particle number $N_i$
with respect to initial conditions. If the minimum goes to zero for
some energy then the probability of the processes $2\to N_f$ is not
exponentially suppressed at 
$E>E_*$~\cite{Rubakov:1995hq}. Otherwise the existence of a nontrivial
minimum at $N_i\sim 1/\lambda$ would indicate on an exponential suppression of
the $2\to N_f$ scattering probability for $E\lsim E_*$. This
idea of exploring the 
classical region was used previously for studies of several different
processes induced by collisions of particles. Among of them are  the false
vacuum decay~\cite{Rubakov:1994hz}, baryon number violating
processes in the Standard Model~\cite{Rebbi:1996zx,Rebbi:1996qt} and 
soliton-antisoliton pair production in the $(1+1)$--dimensional scalar
field theory~\cite{Demidov:2011eu}.

In this paper we study classical solutions describing scattering of the
wave packets in the unbroken $\lambda\phi^4$ theory. In particular,
we try to approach the boundary of the corresponding classically allowed
region.
The motivation for this study is two-fold. First, by identifying
  classically forbidden and allowed regions we can unambiguously
  tell which multiparticle scattering processes are 
  suppressed in the semiclassical limit  and which are not.
  In particular, our results give clear indication for the
  suppression of the probability of $few \to N_f$ processes even in
  the regime $\lambda N_f\gsim 1$, $\lambda\ll 1$ which is in
  agreement with unitarity arguments. Note that it would be very
  interesting to apply the method of classical solutions used in our
  study to explore the classically allowed region in the $(E,N_i,N_f)$
  parameter space in the spontaneously broken $\lambda\phi^4$
  theory in view of direct relevance to the Higgs physics. Another part of our motivation is related to the boundary of the
  classically allowed parameter space and corresponding classical
  solutions. Using suitable semiclassical methods one can start
  obtaining solutions of the classical equations of motion which
  describe multiparticle scattering processes $N_i\to N_f$ in the
  classically forbidden
  region in the semiclassical approximation and calculate the
  semiclassical exponent similar to those in Eq.~\eqref{0_1}. As one
  approaches the boundary of the classically allowed region in the
  $(E,N_i,N_f)$ parameter space the semiclassical exponent is expected
  to nullify and comparison with the boundary and corresponding
  classical solutions 
  obtained in the present study would be a valuable check of the
  semiclassical procedure which is known to plague from multiple
  branches of solutions. 

Classical scattering of waves in relation to multiparticle
production was previously studied to some extent in  the
$(1+3)$-dimensional $\phi^4$ model~\cite{Goldberg:1992es},
$(1+1)$-dimensional abelian Higgs model~\cite{Rajagopal:1991yp} and
non-abelian gauge theories~\cite{Gong:1993bz,Hu:1995ia,Hu:1995kd}.
In particular, authors of Ref.~\cite{Goldberg:1992es} constructed an
initial wave packet consisting of a few high frequency free field
modes and examine its evolution. They found no significant energy
transfer to low frequency modes which would indicate to the production of
many ``quanta''. In our study we use stochastic sampling technique to
scan numerically over classical solutions describing the multiparticle
scattering. We limit ourselves to spherically symmetric solutions
which reduces the system to a $(1+1)$-dimensional model and makes the 
problem numerically feasible.
We find it is technically more convenient to fix not final $N_f$ but 
initial particle number $N_i$ and energy $E$ and find the minimal and
maximal values of the particle number $N_f$ of outgoing wave packets which
is obtained by solving classical equations of motion. Both approaches
are equivalent due to the time reversal symmetry. With our numerical
methods we find a non-trivial classically allowed region for the
multiparticle scattering processes. Namely, we are able to determine
the upper and lower boundaries $N_f^{min}$ and $N_f^{max}$ as
functions of energy $E$ for a set of fixed values of $N_i$. We observe
that the change in particle number in the scattering reaches values up
to 22\% for the studied energy range. We examine properties of the
classical solutions corresponding to the boundaries.

The rest of the paper is organized as follows. Section~2 is devoted to
the description of the model and introduction of notations and most
relevant quantities. In Section~3 we describe numerical method which
we utilize to study solutions to the classical equations of motion
describing scattering of wave packets with  a particular interest to
those ``providing'' maximal change in their particle numbers. In
Section~4 we present our numerical results. Section~5 is reserved for a
discussion and conclusions. 

\section{The model}
We consider $(3+1)$--dimensional model of a scalar field with the
action 
\be
\label{1_1}
S[\phi] = \int d^4x\left[\frac{1}{2}(\partial_\mu\phi)^2 -
\frac{m^2}{2}\phi^2 - \frac{\lambda}{4}\phi^4\right]. 
\ee
The parameter $m^2$ is taken to be positive which corresponds to the
unbroken phase of this theory. By making rescaling $x^\mu\to m^{-1}x^\mu$ and
$\phi\to\sqrt{\frac{m^2}{\lambda}}\phi$ this action can be cast into
\be
\label{1_2}
S[\phi] = \frac{1}{\lambda}\int
d^4x\left[\frac{1}{2}(\partial_\mu\phi)^2 - \frac{1}{2}\phi^2 -
\frac{1}{4}\phi^4\right]. 
\ee
In this form $\lambda$ is a semiclassical parameter which does not
enter equations of motion. We are interested in a particular type of classical
solutions, which describe collisions of wave packets.
In what follows we limit ourselves to
spherically symmetric field configurations. In this case it is useful
to make the following redefinition
\be
\label{1_3}
\phi(t,r) = \frac{1}{r}\chi(t,r)
\ee
and we obtain the action
\begin{equation}
  \label{1_4}
S =\frac{4\pi}{\lambda} \int dt dr \left[\frac{1}{2} \left(
  \frac{\partial \chi}{\partial t} \right)^2 - \frac{1}{2} \left(
  \frac{\partial \chi}{\partial r} \right)^2 - \frac{\chi^2}{2} -
  \frac{\chi^4}{4r^2} \right] 
\end{equation}
of $(1+1)$--dimensional theory on a half-line with the boundary
condition $\chi(t,0)=0$ and the position dependent
interaction. The corresponding equation of motion reads
\be
\label{1_5}
\frac{d^2\chi}{d t^2} - \frac{d^2\chi}{dr^2} + \chi +
\frac{1}{r^2}\chi^3 = 0
\ee
and the energy related to the field configuration $\chi(t,r)$ is
\be
\label{1_6}
E = \frac{4\pi}{\lambda}\int_0^\infty
dr\left[\frac{1}{2}\left(\frac{\partial\chi}{\partial t}\right)^2 +
  \frac{1}{2}\left(\frac{\partial\chi}{\partial r}\right)^2 +
  \frac{1}{2}\chi^2 + \frac{1}{4r^2}\chi^4\right].
\ee
In what follows we solve the classical e.o.m.~\eqref{1_5} numerically.
To do this we restrict our solutions to a finite space interval
$[0,R]$ where $R$ is sufficiently large. 
At $r=R$ we impose Neumann boundary condition
$\partial_r\chi = 0$. This restriction allows us to expand the field
configuration as  follows
\be
\label{1_7}
\chi(t,r) =
\sum_{n=0}^{\infty}c_n(t)\sqrt{\frac{2}{R}}\sin{k_nr},\;\;\;
{\rm where}\;\; k_n=\frac{\pi}{2R}(2n+1),\; n=0,1,...
\ee
 Inserting this expansion
in the action~\eqref{1_4} we obtain the following equations of motion
\be
\label{1_8}
\ddot{c}_n + \omega_n^2c_n + I_n = 0\,,\;\;\; n=0,1,...
\ee
where $\omega_n^2 = k_n^2 + 1$ and the interaction term reads 
\be
\label{1_9}
I_n = \sqrt{\frac{2}{R}}\int_0^{R}dr
\frac{\chi^3(t,r)}{r^2}\sin{k_nr}\,,\;\;\; n=0,1,...
\ee
Total energy of the solution expressed via $c_n(t)$ has the form
\be
\label{1_10}
E = \frac{4\pi}{\lambda}\sum_{n}\left[\frac{1}{2}\dot{c}_n^2 +
\frac{1}{2}\omega_n^2c_n^2 + \frac{1}{4}c_nI_n\right].
\ee
Below we consider solutions which linearize at initial and final
times. In this case they can be interpreted as describing a
multiparticle scattering. In the linear regime the time-dependent Fourier
components $c_n(t)$ can be conveniently written via positive and
negative frequency components as follows
\be
\label{1_11}
c_n(t)\to\frac{1}{\sqrt{2\omega_n}}\left(a_n{\rm e}^{-i\omega_n t} +
a_n^{*}{\rm e}^{i\omega_n t}\right)\;\;\; {\rm as}\;\; t\to -\infty
\ee
and
\be
\label{1_12}
c_n(t)\to\frac{1}{\sqrt{2\omega_n}}\left(b_n{\rm e}^{-i\omega_n t} +
b_n^{*}{\rm e}^{i\omega_n t}\right)\;\;\; {\rm as}\;\; t\to +\infty \tilde{.}
\ee
Using these representations one can compute the energy of colliding
wave packets as
\be
\label{1_13}
E = \frac{4\pi}{\lambda}\sum_n\omega_n\left|a_n\right|^2 =
\frac{4\pi}{\lambda}\sum_n\omega_n\left|b_n\right|^2\;.
\ee
Here $\left|a_n\right|^2$ and $\left|b_n\right|^2$ can be thought as
occupation numbers for initial and final states and their sums
\be
\label{1_14}
N_{i} = \frac{4\pi}{\lambda}\sum_n\left|a_n\right|^2\;,\;\;\;
N_{f} = \frac{4\pi}{\lambda}\sum_n\left|b_n\right|^2\;
\ee
are initial and final particle numbers, respectively. For convenience
we introduce the following shorthand notations
\be
\label{1_15}
\tilde{E} = \frac{\lambda}{4\pi}E\;,\;\;\;
\tilde{N}_i = \frac{\lambda}{4\pi}N_i\;,\;\;\;
\tilde{N}_f = \frac{\lambda}{4\pi}N_f\;.
\ee

For the numerical implementation we truncate the expansion~\eqref{1_7} at
$n=N_r$ and solve the evolution equations~\eqref{1_8} using
Bulirsch-Stoer method, see e.g.~\cite{Press:2007:NRE:1403886}. Some of
the results have been verified with the 4-th order Runge-Kutta method with
a very small time step. For convenience, we introduce the uniform spacial
lattice ${r_i}$, $i=0, ..., N_r$ with $r_0=0$ and $r_{N_r}=R$. In what
follows we use several values of $R=20, 30$ and 
50 and $N_r=400, 600$ and 1000 to study the dependence of our results
on the lattice. The time interval is taken to be somewhat smaller
than $\sim 2R$. We use FFTW implementation of discrete Fourier
transformation~\cite{Frigo05thedesign} to compute the 
field configuration $\chi(t,r)$ as~\eqref{1_7} and the interaction
term~\eqref{1_9}. 

\section{The method}
In this Section we describe a method used to explore the classical
transitions $\tilde{N}_{i}\to \tilde{N}_{f}$ at fixed energy
$\tilde{E}$. In particular, we are interested in solutions of the
classical e.o.m. which maximize $|\tilde{N}_{f}-\tilde{N}_{i}|$ at
fixed values of $\tilde{N}_{i}$ and $\tilde{E}$.  

\subsection{Initial conditions}
\label{sec_3_1}
We take initial conditions which correspond to a wave packet which is
localized well away from the interaction region and which is
propagating towards $r=0$. Technically, we choose an interval
$\left[r_{1},r_{2}\right]$, where $r_1, r_2$ are chosen points
belonging to the spacial lattice, i.e. $r_1=r_{i_1}$ and $r_2=r_{i_2}$
for some $i_i<i_2$, and construct following function
\be
\label{2_1}
\chi(r) =
\begin{cases}
  \sum_{n=1}^{i_2-i_1}\frac{1}{\sqrt{2\tilde{\omega}_n}}f_n\sin{\left(\tilde{k}_n(r-r_{1})\right)}\,
    ,\;\;\; r\in [r_1,r_2]\\
    0, r \notin [r_1,r_2]
\end{cases}
\ee
Here $\tilde{k}_n = \frac{\pi n}{r_{2}-r_{1}}$, $\tilde{\omega}_n
= \sqrt{\tilde{k}_n^2+1}$, $n=1,..,i_2-i_1$ and $f_n$ are initial
Fourier amplitudes. By construction the configuration~\eqref{2_1}
nullifies at the boundaries $r=r_{i_1}$ and $r=r_{i_2}$. Next, to make the
initial configuration smooth enough we introduce an artificial smearing
multiplying~\eqref{2_1} by the function
\be
\label{2_2}
\left({\rm e}^{\frac{r_{i_1}+d-r}{a}}+1\right)^{-1}\cdot\left({\rm e}^{-\frac{r_{i_2}-d-r}{a}}+1\right)^{-1}\;,
\ee
where $d$ and $a$ are chosen parameters. We checked that our results
have a very mild dependence on the precise way of the smearing. In what
follows we fix $d=0.2$ and $a=0.1$ for concreteness. 
Next, we obtain Fourier amplitudes $\tilde{f}_n$ of the smeared
configuration by inverting~\eqref{2_1} and take the initial condition
for the e.o.m.~\eqref{1_8} (i.e. the field configuration and its first time
derivative at initial time $t=t_{i}$) from
\be
\label{2_3}
\chi_{i}(t,r) =
\begin{cases}
  \sum_{n=1}^{i_2-i_1}\frac{1}{\sqrt{2\tilde{\omega}_n}}\tilde{f}_n\sin{\left(\tilde{k}_n(r-r_{1})
    + \tilde{\omega}_n(t-t_{i})\right)}\,
    ,\;\;\; r\in [r_1,r_2]\\
    0, r \notin [r_1,r_2]\,.
\end{cases}
\ee
This corresponds to the initial wave packet propagating towards
$r=0$. 
In what follows we set $t_i=0$. We consider only solutions whose
evolution is linear at initial (and 
final) times.

\subsection{Going to the classical boundary}
Using the initial field configuration we calculate its energy $\tilde{E}$ and
initial particle number $\tilde{N}_i$ by using
Eqs. ~\eqref{1_10},~\eqref{1_11},~\eqref{1_13} and~\eqref{1_14}.   
Our aim is to find the classical solutions describing scattering of
wave packets in which the particle number changes as much as possible at a
given energy $\tilde{E}$. Since we solve initial value problem it is
more convenient to fix the initial particle number $\tilde{N}_i$ and find the
absolute minimum 
(or maximum) of $\tilde{N}_{f}$ with respect to chosen set of initial conditions,
i.e. $f_n$ entering~\eqref{2_1}. We fix the initial particle number by the respectful normalization of
the Fourier amplitudes $\tilde{f}_n$.  The final particle number $\tilde{N}_{f}$ is a
highly non-linear function of initial data $\tilde{f}_n$ and it may
have several local extrema. To find its absolute minimum (maximum) we
use the stochastic sampling technique in combination with the simulated
annealing method, see~\cite{Metropolis1953,Press:2007:NRE:1403886}.
Namely, we generate an ensemble of the classical solutions with a fixed initial
particle number $\tilde{N}_{i}$ weighted by the probability
\be
\label{3_3}
p\sim {\rm e}^{-F}\;,
\ee
where
\be
\label{3_4}
F = \beta\left(\tilde{N}_{f} + \xi(\tilde{E}-\tilde{E}_*)^2\right).
\ee
If large positive $\beta$ and $\xi$ are taken the ensemble will be
dominated by solutions having small $F$ thus driving their
distribution towards the lower boundary $\tilde{N}_{f}^{min}(\tilde{E}_*)$
of the classically allowed region. To reach the upper boundary
$\tilde{N}_{f}^{max}(\tilde{E}_*)$ signs of $\beta$ and $\xi$ should
be negative. 

To generate the ensemble~\eqref{3_3} we use the Metropolis Monte Carlo
algorithm. We start with a solution specifying by a randomly chosen
set of $f_n$, see~\eqref{2_1}, which have a fixed initial particle number
$\tilde{N}_i$. This  
solution has an energy $\tilde{E}$ and final particle number
$\tilde{N}_{f}$ which is found after solving the classical equations of
motion. Next, we choose a few randomly chosen amplitude numbers and
change corresponding amplitudes $f_n$ by small amounts, $f_n\to f_n^{\prime}
= f_{n} + \Delta f_n$. We used to take three
amplitudes at a time which shows a good performance of our algorithm. 
The quantities $\Delta f_n$ are chosen to be normally distributed with
the standard deviation $a/\sqrt{\tilde{\omega}_n}$, where $a$ is a
small number. In our calculations its value is varied from $10^{-4}$ to
0.1. The modified set of amplitudes $f_n^\prime$ is rescaled in a proper
way and used to construct an initial wave packet having the same initial
particle number $\tilde{N}_{i}$ 
and some value of energy $\tilde{E}^\prime$, see Section~\ref{sec_3_1}. Then
we evolve the system forward in time far enough until it reaches the
linear regime and we can find the final particle number
$\tilde{N}_{f}^\prime$ of the new solution. We compute $\Delta F\equiv F^\prime - F$
using~\eqref{3_4}. The new set of amplitudes $f_n^\prime$ is accepted
with probability
\be
\label{3_5}
p_{accept} = {\rm min}\left(1,{\rm e}^{-\Delta F}\right)
\ee
and used for the next iteration. Typically, we fix $\tilde{N}_{i}$ and $\tilde{E}_*$ and
perform several runs starting with different $\beta$ and
$\xi$. Each run spans of order $10^3-10^4$ iterations. The value of
$\xi$ remains a constant during a single run, while $\beta$ which is
analogue of inverse temperature is gradually 
increased from its initial value $\beta_0$ according~\cite{4767596} to 
\be
\label{3_6}
\beta_i = \beta_0\log{(1+i)}\;,
\ee
where $i$ is the iteration number. Values of $\beta_0$ and $\xi$ are
temporarily increased from 10 to $10^6$ and from $10^{-4}$ to about 1,
respectively. We stop the procedure when relative change in the value
of $F$ during a run becomes smaller than $10^{-3}$ for the same $\xi$
and it does not increases with the variation of $a$ and $\beta_0$.

Let us note, that the stochastic sampling technique
was previously used to study classically allowed sphaleron
transitions~\cite{Rebbi:1996zx,Rebbi:1996qt} and classical
soliton-antisoliton pair production in particle collisions in 
(1+1)-dimensional scalar field theory~\cite{Demidov:2011eu}, where
its efficiency in determination of the boundary of the classically allowed
region was confirmed. 

\section{Numerical results}
\begin{figure}[!t]
  \begin{center}
    \includegraphics[height=8.0cm]{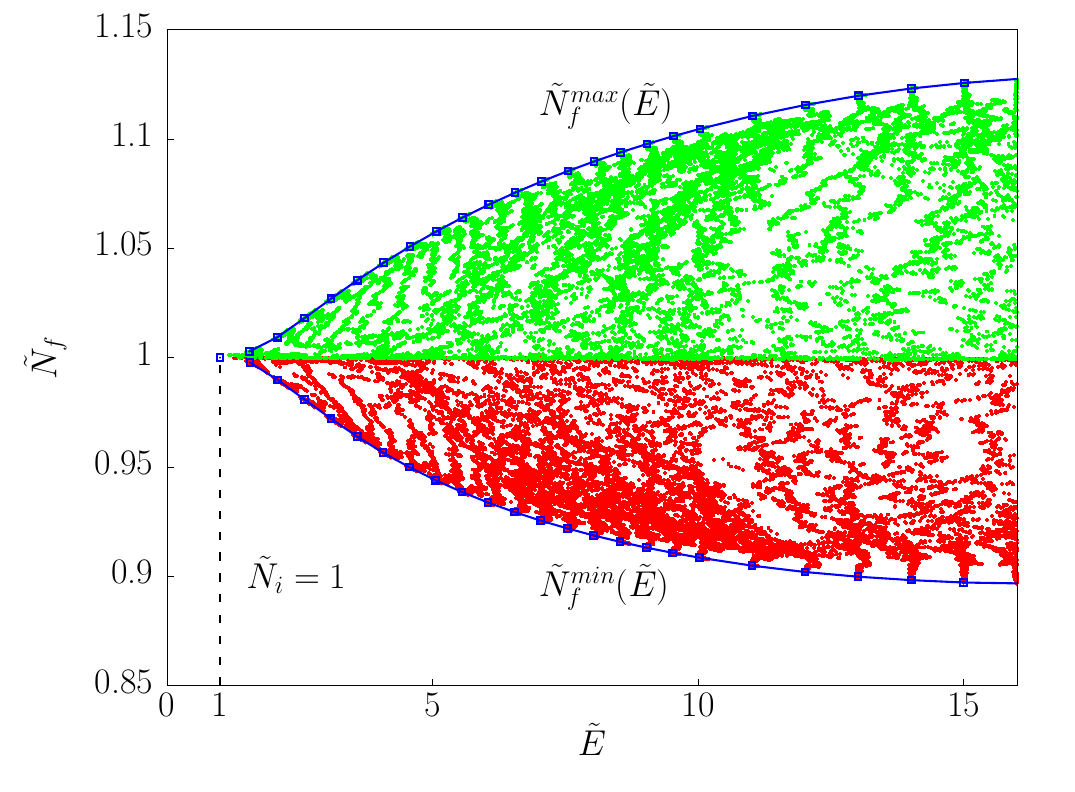}
  \end{center}
\caption{The results of stochastic sampling for $\tilde{N}_i=1$, $N_r=400$
  and $R=20$. Each dot in the figure represents a classical solution
  with an energy $\tilde{E}$ and final particle number $\tilde{N}_f$
  obtained using the Monte-Carlo technique described in the
  text.  Red (dark) and green (gray) points correspond to positive and
  negative values of $\beta$ and $\xi$, respectively. The dashed line
  marks the threshold energy $\tilde{E}=\tilde{N}_f=\tilde{N}_i\equiv
  1$. The upper and lower boundaries, $\tilde{N}_f^{max}(\tilde{E})$ and
  $\tilde{N}_f^{min}(\tilde{E})$ are shown by solid lines. \label{fig_N1_Nx400_all}}
\end{figure}

Below we describe regions of the classical solutions for several
different values of $\tilde{N}_{i}=0.1, 1.0, 10.0$ and 30.0. 
In Fig.~\ref{fig_N1_Nx400_all} 
we plot numerical results of the stochastic sampling in the
$(\tilde{E},\tilde{N}_f)$ plane for $\tilde{N}_i=1$, $N_r=400$ and
$R=20$. The initial field configurations are set in the interval 
$[r_1, r_2]\equiv [5.0,19.0]$. Each dot in the Figure represents a
classical solution accepted by our numerical procedure on the way to
the boundaries 
$\tilde{N}_f^{max}(\tilde{E})$ and $\tilde{N}_f^{min}(\tilde{E})$.
To reach the boundaries we run the numerical procedure described in
the previous Section for a chosen
set of $\tilde{E}_*$ 
\be
\label{4_1}
\frac{\tilde{E}_*}{\tilde{N}_i} = 1.5, 2.0, ..., 9.5, 10.0, 11.0,
..., 15.0\;,
\ee
which determines the relevant energy interval for our study. We
comment on this choice below. The
solutions which seed our Monte Carlo search are randomly chosen with
the the only condition that they have $\tilde{N}_i=1$ and linearize at the
initial and final times. These solutions typically have very small 
change in the particle number  and are situated near the line $\tilde{N}_f=
1$ on the $(\tilde{E},\tilde{N}_f)$ plane.
 Initially, we take it relatively large
value of $a=0.1$ and start our stochastic 
sampling with small $\beta_0$ about 10.0. During single run its value
is changed according to~\eqref{3_6}. We repeat the runs with a larger
starting value of $\beta$. The value of $a$ which determines size of
the Fourier amplitude changes $\Delta f_{n}$ is also decreased after
several runs. At the 
same time the value of $\xi$ is increased to fix the energy of the
solution to be near $\tilde{E}_*$.
The obtained domain of the classical solutions in
$(\tilde{E},\tilde{N}_f)$ plane, see Fig.~\ref{fig_N1_Nx400_all}, has
smooth envelopes, $\tilde{N}_f^{min}(E)$ and $\tilde{N}_f^{max}(E)$, which represent
the boundary of the classically allowed region of the lattice  
version of the continuous model.   We see that the maximal possible
change in the particle number $|\tilde{N}_f-\tilde{N}_i|$ during the
classical evolution  increases with the energy of solutions but does
not exceed 12\% in the chosen energy range. Let us stress that due to the
time reversal symmetry one can always interchange the initial
$\tilde{N}_i$ and final $\tilde{N}_f$ particle numbers in the
discussion of the classically allowed region in the Fig.~\ref{fig_N1_Nx400_all}. 

\begin{figure}[!htb]
  \begin{center}
    \includegraphics[height=8.0cm]{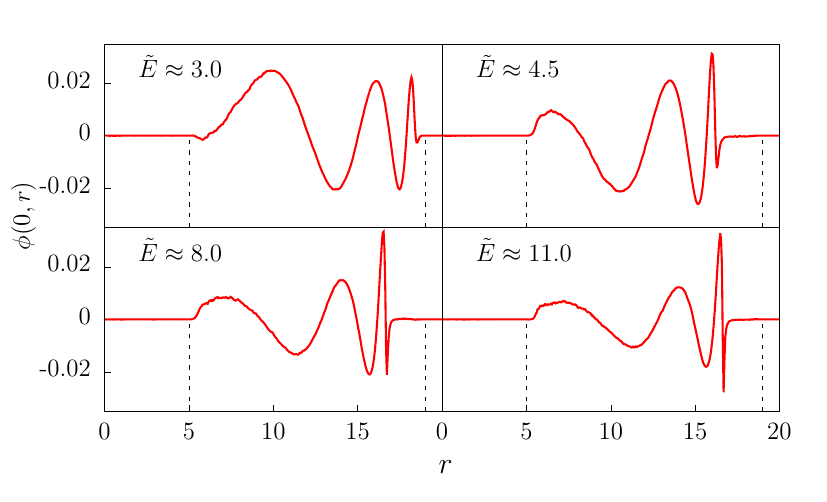}
  \end{center}
\caption{The field configurations $\phi(t=0,r)$ of the solutions at
  the upper boundary with $\tilde{N}_f = \tilde{N}^{max}_f(\tilde{E})$ for
  $\tilde{E}\approx3.0, 4.5, 8.0$ and 11.0. The space lattice
  parameters are   $N_r=400$, $R=30$. The dashed lines mark the space
  interval for the initial wave packet.\label{fig_examples_N1}}
\end{figure}

\begin{figure}[!t]
  \begin{center}
    \includegraphics[height=8.0cm]{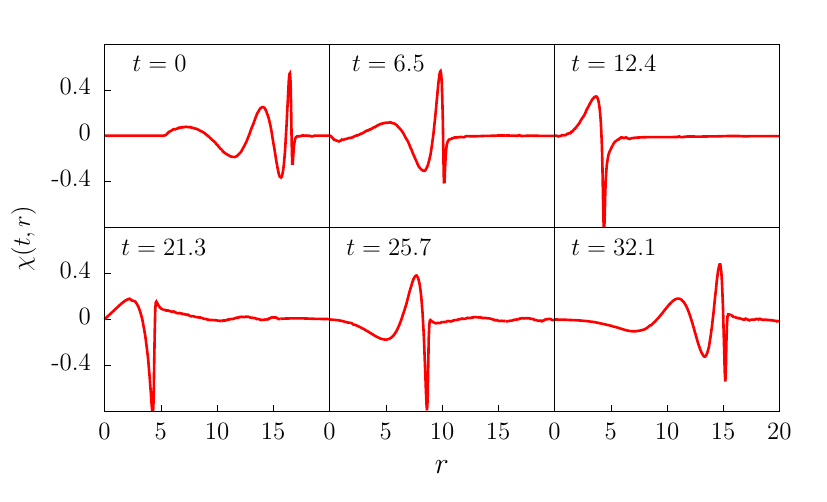}
  \end{center}
\caption{Time evolution of the field $\chi(t,r)=\frac{\phi(t,r)}{r}$ for the upper
  boundary solution with $\tilde{E}\approx6.1$; $N_r=400, R=20$.\label{fig_example_N1}}
\end{figure}

Let us study properties of the classical solutions near the
boundaries $\tilde{N}_f^{min}(E)$ and $\tilde{N}_f^{max}(E)$. Although
the solutions which seed our Monte Carlo runs for each value of
$\tilde{E}_*$ are chosen at random it is remarkable that the boundary
solutions have similar forms for different $\tilde{E}_*$. At the
panels of Fig.~\ref{fig_examples_N1} 
we plot initial wave packets of the classical solutions at the upper
boundary $\tilde{N}_{f}=\tilde{N}_{f}(\tilde{E})$ for several values
of energy $\tilde{E}\approx3.0, 4.5, 8.0$ and~11.0. The interval
$[r_1,r_2]$ is marked by the dashed lines. Corresponding final
particle number for these solutions can be found in the
Fig.~\ref{fig_N1_Nx400_all}. The evolution in time of the
upper boundary solution with $\tilde{E}\approx 6.1$ is presented in
Fig.~\ref{fig_example_N1}
where $\chi(t,r)=\frac{\phi(t,r)}{r}$ is shown for convenience. 
For all boundary solutions which we have found with $\tilde{N}_i=1$ the
initial wave packet has a relatively sharp part near its end. This
part becomes more pronounced with increase of its energy as seen in
Fig.~\ref{fig_example_N1}. The initial wave packet propagates towards the
interaction region and its spiky parts increases considerably near the
point $r=0$. The reflected wave packet have the form very similar to 
the incoming one. The spiky part of the incoming wave packet is the
last to arrive into the interaction region and the first to leave it.

\begin{figure}[!htb]
  \begin{center}
    \begin{tabular}{cc}
      \includegraphics[height=5.0cm]{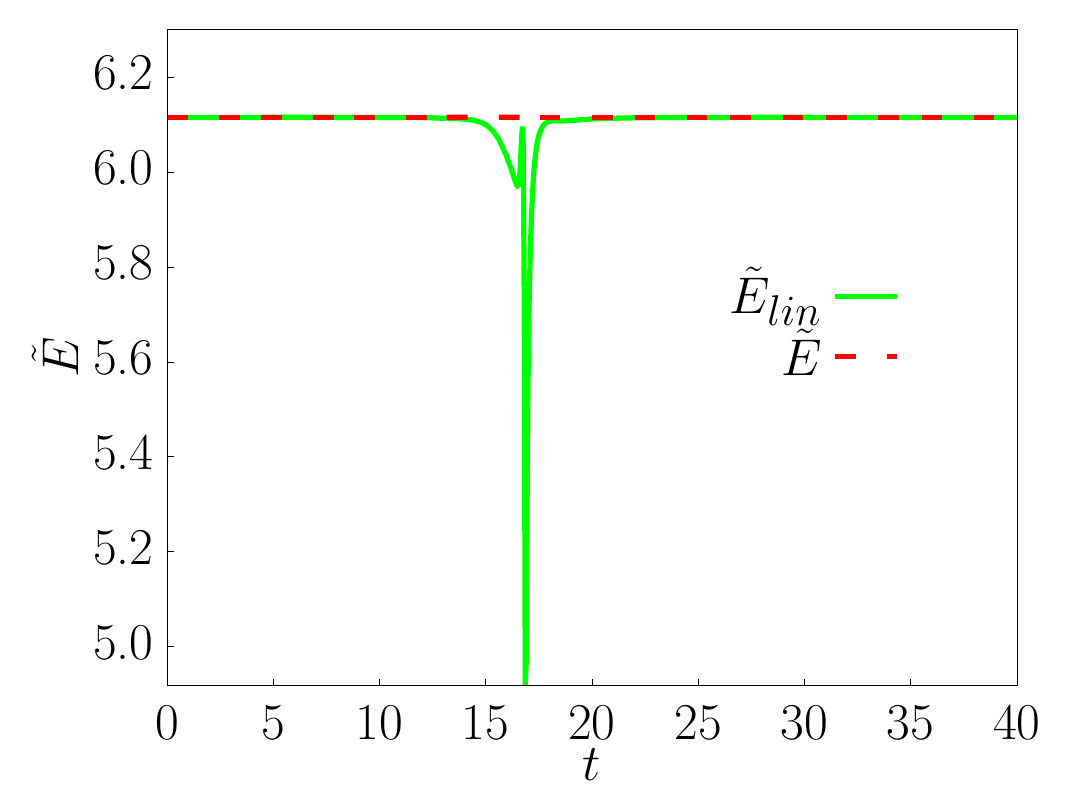}&
      \includegraphics[height=5.0cm]{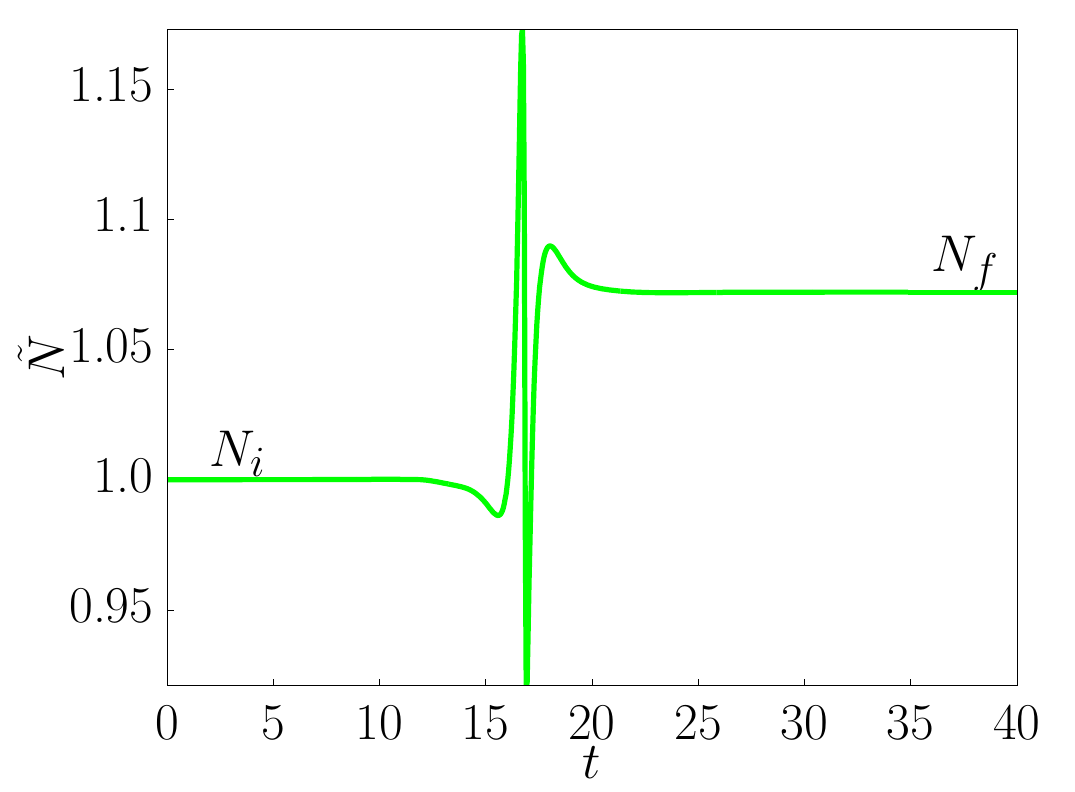}
      \end{tabular}
  \end{center}
\caption{Time evolution of the linearized and exact energy (left panel) as
  well as the instantaneous particle number (right
  panel) for the exemplary solution in
  Fig.~\ref{fig_example_N1}. \label{fig_EN_example_N1}}  
\end{figure}

According to the used extremization procedure which
involves the stochastic sampling, the found classical solutions describe the
scattering of waves with $\tilde{N}_i=1$ and a maximum value of
$|\tilde{N}_f-\tilde{N}_i|$ at a given energy $\tilde{E}$. It is
interesting  to study how the particle number changes during the time 
evolution. In Section~2 we introduced the particle numbers for 
  the initial and final wave 
packets. However, we can calculate an instantaneous particle number
$N(t)$ by using Eq.~\eqref{1_11} (or~\eqref{1_12}) as a definition of
time-dependent positive and negative frequency components at arbitrary
time $t$ and applying
formulas similar to~\eqref{1_14}. The quantity
$\tilde{N}(t)$ coincides with $\tilde{N}_i$ and $\tilde{N}_f$ at
initial and final times where the evolution of the field is linear.
On the right panel of the
Fig.~\ref{fig_EN_example_N1} we plot the instantaneous particle
number for the boundary solution shown in
Fig.~\ref{fig_example_N1}. Comparing it with the 
time evolution of the field presented on Fig.~\ref{fig_example_N1} we
see that the actual change of the particle number occurs precisely
when the sharpest part of the initial wave packet reaches the
interaction region. For instance, the deviations of $\tilde{N}$ at
time stamps $t=12.4$ and $t=21.3$ (see Fig.~\ref{fig_example_N1}) from
its asymptotic values $\tilde{N}_i$ and $\tilde{N}_f$ are less than
0.0005, which are much smaller then $\tilde{N}_f-\tilde{N}_i\approx
0.072$. Such behaviour of $\tilde{N}(t)$ give us support that found
extrema of $F$ are independent of choice of the interval $[r_1,r_2]$
which is used to generate initial field configurations as far as it is
sufficiently large and include the part of the wave packet responsible
for the most of the change in $\tilde{N}(t)$. Also we verified that 
numerical error in $|\tilde{N}_f-\tilde{N}_i|$ related to the smearing is
less than $10^{-3}$. In a way similar to
$\tilde{N}(t)$ we define a linearized energy $\tilde{E}_{lin}(t)$ 
which is used to check the linearity of the solution by comparing it
with the exact energy~\eqref{1_10}. Such comparison is presented in
Fig.~\ref{fig_EN_example_N1} (left panel)
for the solution from Fig.~\ref{fig_example_N1}. The total energy is
conserved on the entire solution with the accuracy less than
$10^{-3}$, while it coincides with the linearized energy better than
$10^{-4}$ at initial and final times.

\begin{figure}[!htb]
  \begin{center}
    \includegraphics[height=8.0cm]{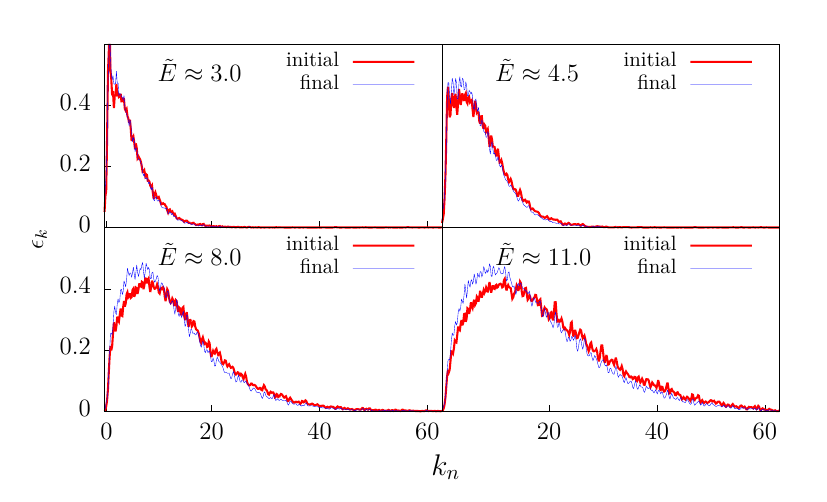}
  \end{center}
\caption{The distributions of energy over the modes $\epsilon_k$ per wave
  number for the
  initial (red, thick line) and final (blue, thin line) wave
  packet for the boundary solutions from
  Fig.~\ref{fig_examples_N1}.\label{fig_ek_example_N1}}  
\end{figure}

A nontrivial change in particle number implies a redistribution of the
energy between low and high frequency modes. In
Fig.~\ref{fig_ek_example_N1} 
we show energy distributions $\epsilon_k$ per wave number unit for
initial and final wave packets for the same upper boundary solutions in 
Fig.~\ref{fig_examples_N1}. These distributions are defined as
\be
\label{4_2}
\epsilon_k =
\begin{cases}
  \frac{1}{\Delta k}\omega_n\left|a_n\right|^2\,, \;\;\; {\rm for}\; {\rm the}\; {\rm initial}\; {\rm wave}\;
        {\rm packet}\,,\\
  \frac{1}{\Delta k}\omega_n\left|b_n\right|^2\,, \;\;\; {\rm for}\; {\rm the}\; {\rm final}\; {\rm wave}\; {\rm
    packet}\,,
\end{cases}
\ee
where $\Delta k = \frac{\pi}{R}$. We observe expected softening of the 
energy distributions for the final wave packet as compared to the
initial one. However, the effect is quite small even for boundary
solutions which describe processes with maximal change of  the particle
number. Also we see that with increase of the collision energy $\tilde{E}$
the modes with larger $k_n$ (i.e. higher frequencies) become filled in and at
$\tilde{E}\gsim 11.0$ the tail of energy distributions reaches the
largest value of $k_n$ for the chosen lattice size and spacing. This is a
clear indication that a finer lattice is required to obtain reliable
results at these and higher energies.
At the same time solutions with
energies 
not far from the threshold energy $\tilde{E}_{th}=\tilde{N}_i$
consist mostly of nonrelativistic modes and a larger space interval $R$
is needed to reach linear regime at initial and final times. For this
reason we do not consider solutions with energies lower than 1.5.

\begin{figure}[!htb]
  \begin{center}
    \includegraphics[height=8.0cm]{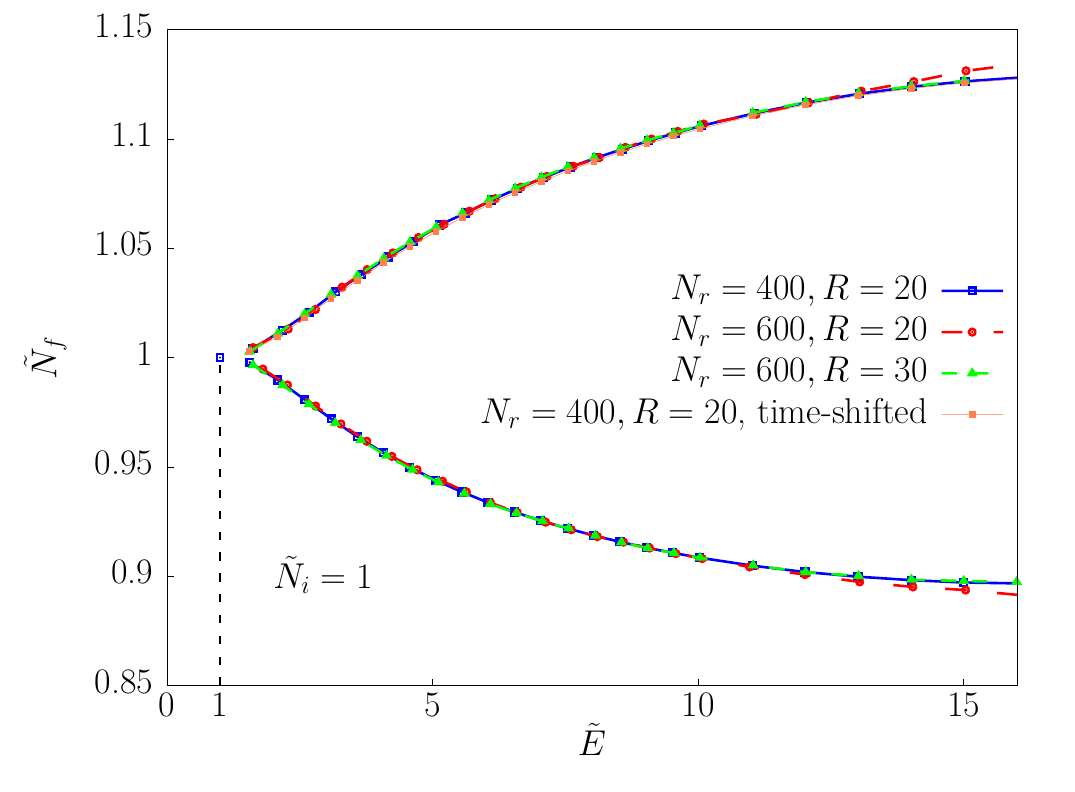}
  \end{center}
\caption{The classically allowed region in $(\tilde{E}, \tilde{N}_f)$
  plane for $\tilde{N}_i=1$ calculated for three cases: 1) $N_r=400,
  R=20$; 2) $N_r=600, R=20$; 3) $N_r=600, R=30$; 4) $N_r=400, R=30$,
  time-shifted solutions (see main text for details) . The space intervals
  chosen for selection of the initial wave packets are presented in
  the main text.\label{fig_N1_Nx400_lattice}} 
\end{figure}

\begin{figure}[!htb]
  \begin{center}
    \includegraphics[height=8.0cm]{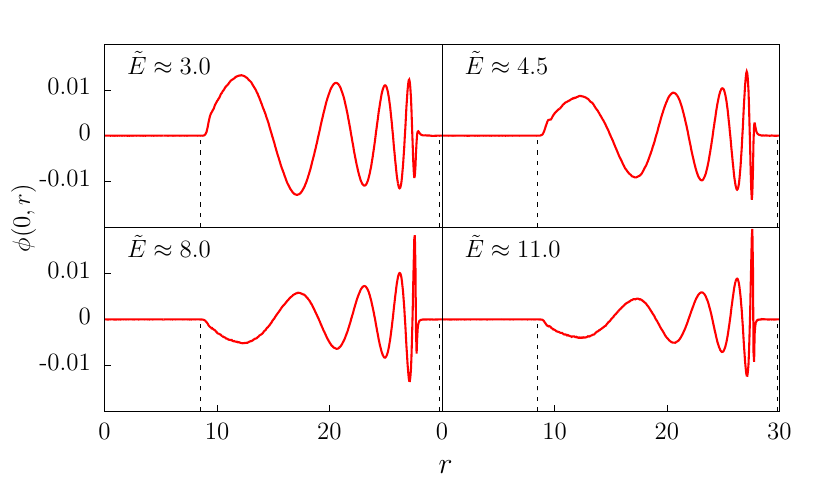}
  \end{center}
  \caption{The same as in Fig.~\ref{fig_examples_N1} but for $N_r=600,
    R=30$.\label{fig_examples_N1_L30}}
\end{figure}

\begin{figure}[!htb]
  \begin{center}
    \includegraphics[height=8.0cm]{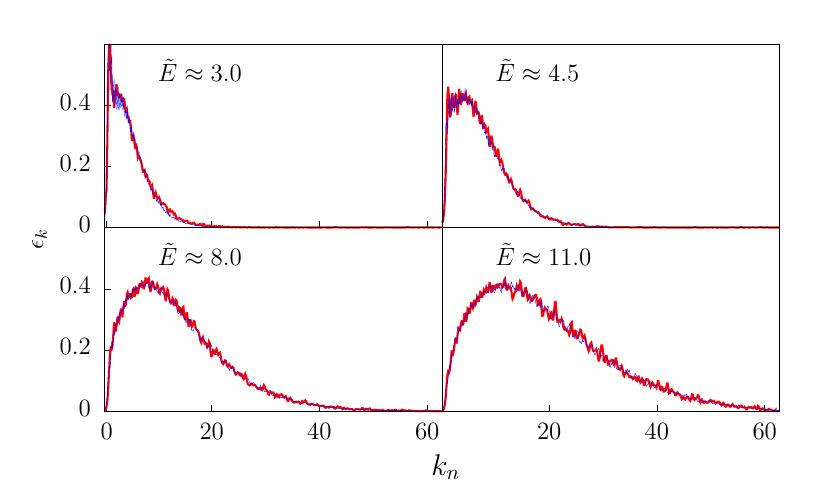}
  \end{center}
\caption{The distribution of energy over the modes $\epsilon_k$ per wave
  number for
  the upper boundary solutions for $N_r=400, R=20$ (red, thick line)
  and $N_r=600, R=30$ (blue, thin line)   Figs.~\ref{fig_examples_N1}
  and~~\ref{fig_examples_N1_L30}.\label{fig_ek_example_N1_L30}}   
\end{figure}

Next, we study the dependence of the obtained results for the boundary of
the classically allowed region in the $(\tilde{E}, \tilde{N}_f)$
plane for $\tilde{N}_i=1$ and for the boundary solutions on parameters of the lattice, namely on the
spacial cutoff $R$ and on the number of modes $N_r$. For comparison to the
case with $R=20$ and $N_r=400$ we repeat the same numerical procedure
to find the boundary of the classically allowed region for the case $R=20,
N_r=600$ with a smaller lattice spacing and for the case $R=30, N_r=600$
with a larger space interval but the same lattice spacing. The results
are presented in 
Fig.~\ref{fig_N1_Nx400_lattice}, 
where we show lower and upper boundaries $\tilde{N}_f^{min}(E)$ and
$\tilde{N}_f^{max}(E)$ for the different cases. The space intervals for selection of the
initial wave packets are taken to be $[6.7,19.2]$ in the former case
and $[8.5,29.8]$ in the latter.  
We observe that different boundaries coincide with accuracy better
than $4\cdot 10^{-3}$. A deviation is
seen for the case $N_r=600, R=30$ for large energies $\tilde{E}\gsim
11$ which reflects the appearance of new high frequency modes. 
Examples of the upper boundary solutions for the same set of energies
as in Fig.~\ref{fig_examples_N1} but for the case $N_r=600, R=30$ are
shown in Fig.~\ref{fig_examples_N1_L30}.
Comparing them with the solutions presented in
Fig.~\ref{fig_examples_N1} we see that the difference lies in the enlarged  
soft oscillating part of the solution. We found that the time
evolution of the particle number $\tilde{N}(t)$ for these  solution is
similar to that of presented in Fig.~\ref{fig_EN_example_N1}. Namely,
the actual change in $\tilde{N}$ starts with arrival of the sharp
tail of the incoming wave train into the interaction region. 
Energy distributions over Fourier modes for the solutions with the
same energy obtained on different lattices  are almost
indistinguishable as seen from Fig.~\ref{fig_ek_example_N1_L30}.   

\begin{figure}[!htb]
  \begin{center}
    \includegraphics[height=8.0cm]{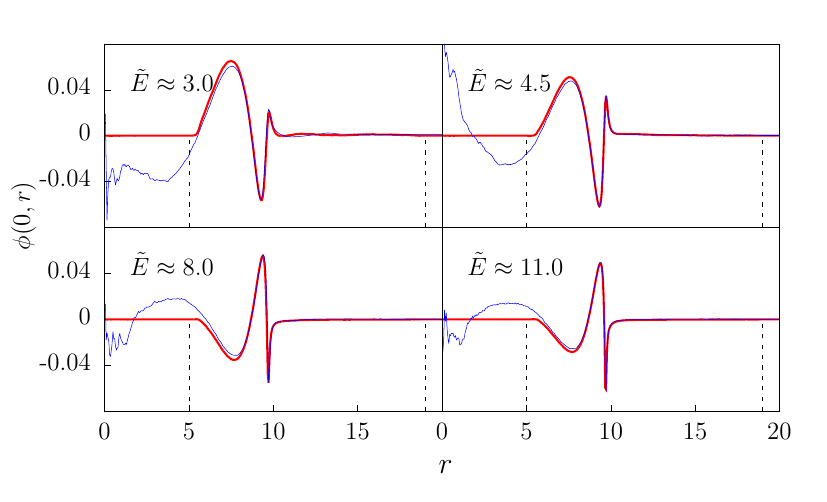}
  \end{center}
\caption{The upper boundary initial field configurations of the solutions
  related by time-shift symmetry for   $\tilde{E}\approx3.0, 4.5, 8.0$
  and 11.0. The space lattice parameters are   $N_r=400$, $R=20$. The
  dashed lines mark the space 
  interval for the initial wave packet.\label{fig_examples_N1_shift}}
\end{figure}

Our numerical results clearly show existence of the classically
allowed region of a finite size in the $(\tilde{E},\tilde{N}_f)$ plane for
fixed value of $\tilde{N}_i$ in the continuous version of the model
with the $\phi^4$ potential.  We suggest existence of the 
boundary solutions in the continuous limit of the model.
The classical equations of motion~\eqref{1_5} and~\eqref{1_8} are
invariant with respect to time translations. This symmetry is
explicitly broken in our
numerical setup by the choice of the initial configuration~\eqref{2_1}
and~\eqref{2_3} at $t=t_i\equiv 0$, which in particular implies
\be
\label{4_3}
\chi(0,r_1)=\chi(0,r_2)=0\,.
\ee
Still the oscillating form of the initial wave packets (see
Figs.~\ref{fig_examples_N1} and~\ref{fig_examples_N1_L30}) and
the time evolution of the particle number indicate that an
approximate discrete time-shift symmetry should exists in the lattice
version of the model. In terms of initial wave packets  this
symmetry connects the configurations which are related by time
evolution and which satisfy conditions~\eqref{4_3} as long as the
spiky part of the initial wave train lies inside the interval
$[r_1,r_2]$ .  Indeed, with our numerical technique (making several
runs at the same energy $\tilde{E}_*$ with different seeds) we
actually find several branches of the boundary solutions which are related with each
other by such shifts in time. In fact, for constructing the boundaries in
Fig.~\ref{fig_N1_Nx400_all}  we selected the solutions whose initial
wave packets have the longest soft oscillating part while their sharp
edges are placed as far as possible (for given $R$ and initial space
interval $[r_1,r_2]$) from the origin.  
In Fig.~\ref{fig_examples_N1_shift}      
we plot the initial field configurations (red solid lines) corresponding
to another branch of the upper boundary solutions at the same values
of energy as in Fig.~\ref{fig_examples_N1}. These wavepackets have
spiky parts shifted closer to the origin as compared to those in
Fig.~\ref{fig_examples_N1}. Also in the
Fig.~\ref{fig_examples_N1_shift} in thin (blue) lines we show the
initial field configurations from Fig.~\ref{fig_examples_N1} evolved
forward in time until the positions  of the
spiky parts of the different solutions coincide. We observe perfect
coincidence of the form of that part. As the spiky part of the
wavepackets is responsible
for the change of the particle number we expect that time-shifted
solutions would have similar values of $\tilde{N}_f$. This is indeed
the case and in Fig.~\ref{fig_N1_Nx400_lattice} we show the upper
boundary $\tilde{N}_t^{max}$ obtained for $N_{r}=400$ and $R=20$ with
the branch of the time shifted solutions. It coincides with that of
for the upper boundary solutions with a longer soft
oscillating part with an accuracy better than $2\cdot 10^{-3}$.  

\begin{figure}[!htb]
  \begin{center}
    \includegraphics[height=8.0cm]{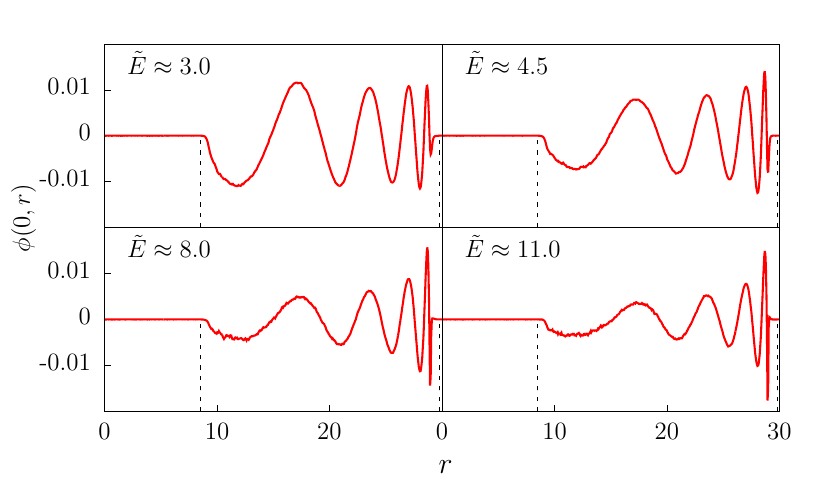}
  \end{center}
  \caption{The same as in Fig.~\ref{fig_examples_N1_L30} but for lower
    boundary solutions.\label{fig_examples_N1_L30_down}}
\end{figure}

So far we described the classical solutions at the upper part
$\tilde{N}_f=\tilde{N}_f^{max}(E)$ of the boundary.  Examples of solutions
corresponding to the lower part $\tilde{N}_f=\tilde{N}_f^{min}(E)$ of
the boundary are shown in Fig.~\ref{fig_examples_N1_L30_down}
for the case $N_r=600, R=30$. We observe that they have very similar 
properties.  Let us note that the time reversal symmetry translates a
lower boundary solution with initial $\tilde{N}_i$ and final
$\tilde{N}_f$ particle numbers into an upper boundary solution with
the initial and final particle numbers interchanged.

\begin{figure}[!htb]
  \begin{center}
    \includegraphics[height=8.0cm]{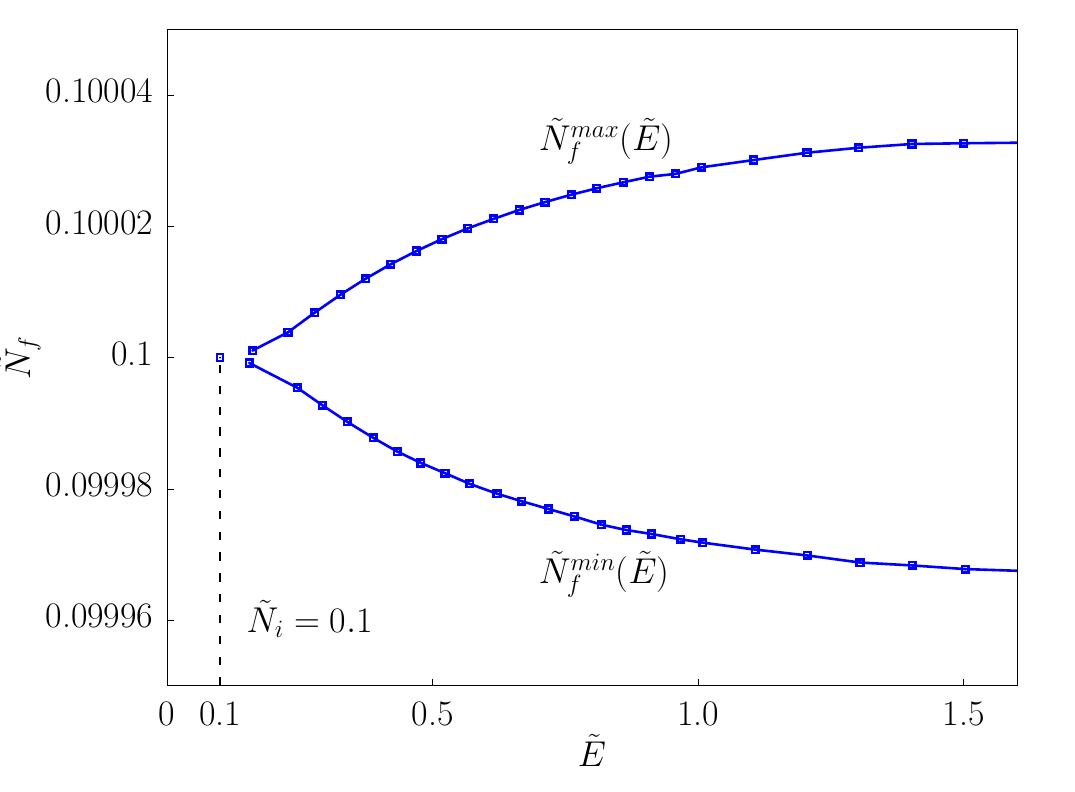}
  \end{center}
\caption{ The classically allowed region in $(\tilde{E}, \tilde{N}_f)$
  plane for $\tilde{N}_i=0.1$; $N_r=600, R=30$. \label{fig_N0_1}} 
\end{figure}

\begin{figure}[!htb]
\begin{picture}(300,240)(0,20)
\put(220,20){\includegraphics[angle=0,width=0.47\textwidth]{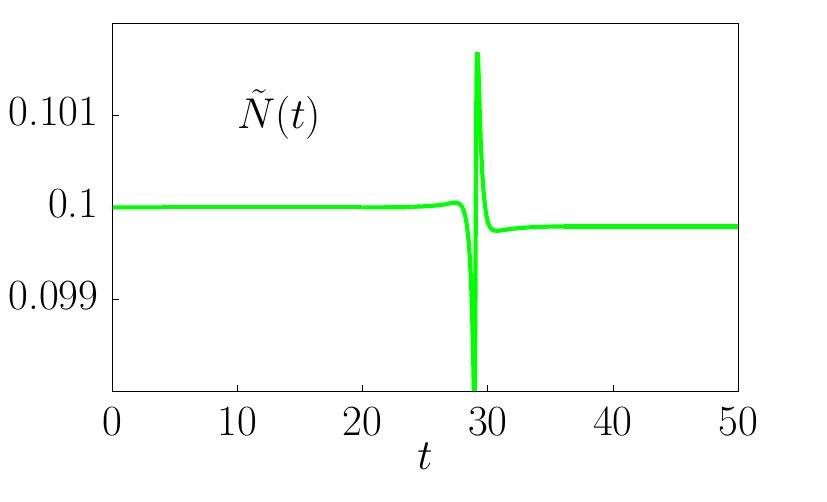}}
\put(20,20){\includegraphics[angle=0,width=0.47\textwidth]{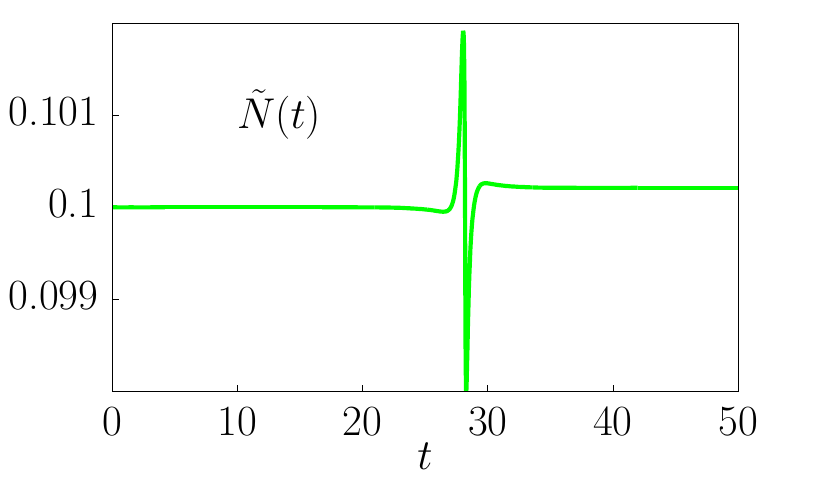}}
\put(220,140){\includegraphics[angle=0,width=0.47\textwidth]{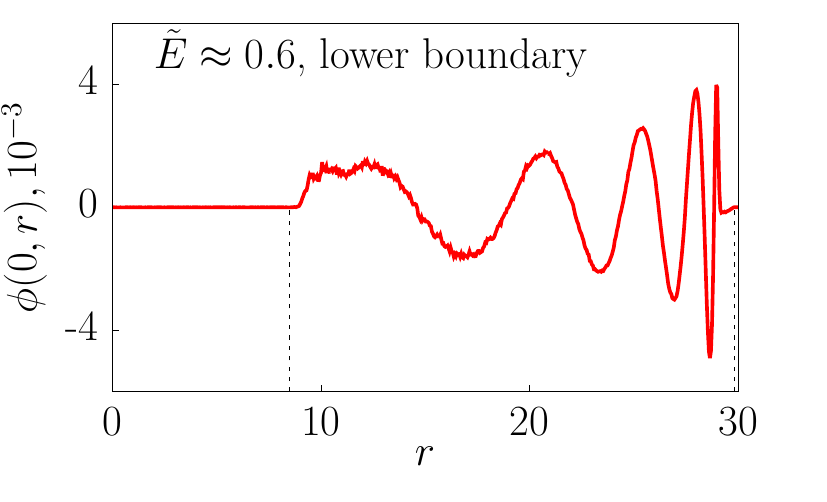}}
\put(20,140){\includegraphics[angle=0,width=0.47\textwidth]{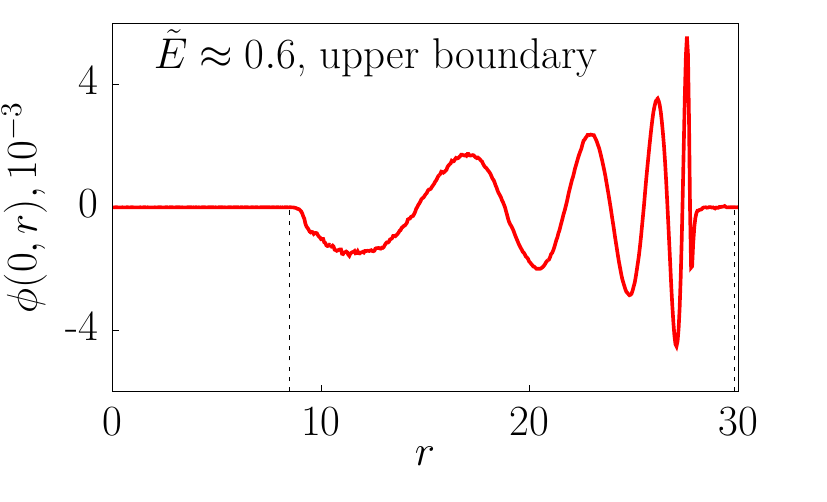}}
\end{picture}
\caption{\label{fig_examples_N0_1} The initial wave packets (left panels) and 
  evolution of the instantaneous particle number (right panels) for
  boundary solutions with $\tilde{N}_i=0.1$ and $\tilde{E}\approx 0.6$. Upper
and lower panels correspond to upper and lower boundary solutions
respectively. }
\end{figure}

Now we turn to discussion of numerical results for another values 
of the initial particle number $\tilde{N}_i$.  In Fig.~\ref{fig_N0_1}
we show the upper $\tilde{N}_f(E)$ and lower $\tilde{N}_f(E)$
boundaries of the classically allowed region for $\tilde{N}_i=0.1$ and the
relevant energy interval
$\left[1.5\tilde{N}_i,15.0\tilde{N}_i\right]$ with the values of
$\tilde{E}_*$ defined by~\eqref{4_1}. We see that  the maximum  
relative difference of the initial and final particle 
numbers, i.e. $\frac{|\tilde{N}_f-\tilde{N}_i|}{\tilde{N}_i}$,  is
more than two order of magnitude smaller than for the case
$\tilde{N}_i=1$ for energies $\tilde{E}\lsim 15\tilde{N}_i$.
Numerically, the particle number changes no more than 0.33\% for the
chosen energy range. In Fig.~\ref{fig_examples_N0_1}
we show examples of the initial wave packets and instantaneous particle
number evolution for the upper and lower boundary solutions for
$\tilde{E}\approx0.6$. The space interval for the initial wavepacket
$[r_1,r_2]$ is shown by the dashed lines. Qualitative properties of these solutions are very
similar to those obtained for $\tilde{N}_i=1$. Namely, the solutions
have a spiky part whose transition through the interaction region results
in actual change in particle number and a soft oscillating part. As in
the previous case we obtain several branches of the boundary solutions
which differ by corresponding shifts in time. 

\begin{figure}[!htb]
  \begin{center}
    \includegraphics[height=8.0cm]{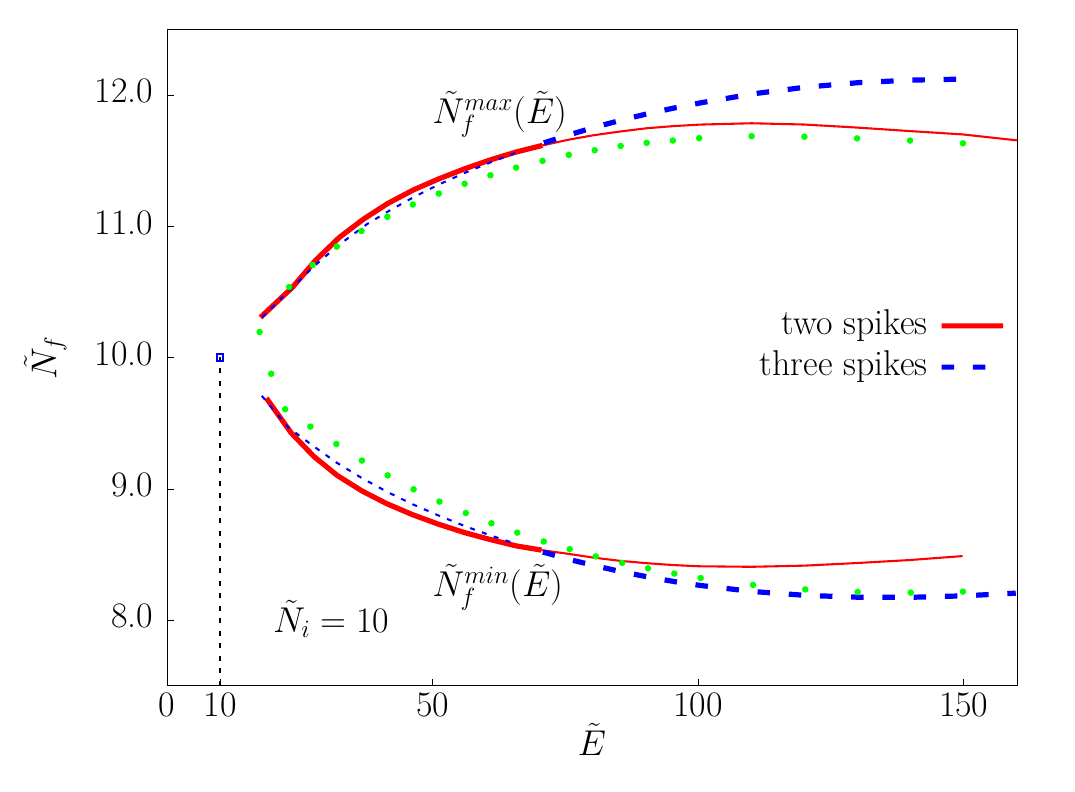}
  \end{center}
\caption{The classically allowed region in $(\tilde{E}, \tilde{N}_f)$
plane for $\tilde{N}_i=10.0$; $N_r=600, R=30$. The envelopes
$\tilde{N}_{f}^{min}$ and $\tilde{N}_{f}^{max}$ are composed of the
classical solutions with two spiky parts (thick red line) and with
three spiky parts (dashed blue line). The dots show other branches of the solutions which deliver some local minima to $F$ (or $\tilde{N}_{f}$).
  \label{fig_N10}} 
\end{figure}

\begin{figure}[!htb]
\begin{picture}(300,480)(0,20)
\put(220,20){\includegraphics[angle=0,width=0.47\textwidth]{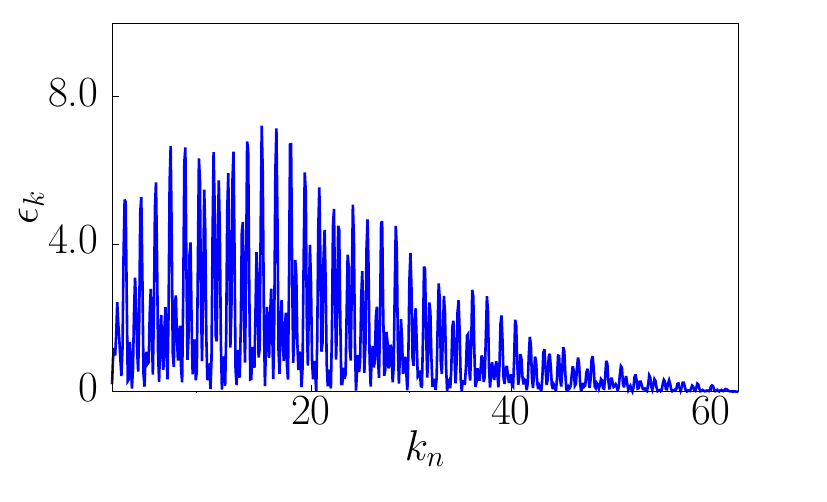}}
\put(20,20){\includegraphics[angle=0,width=0.47\textwidth]{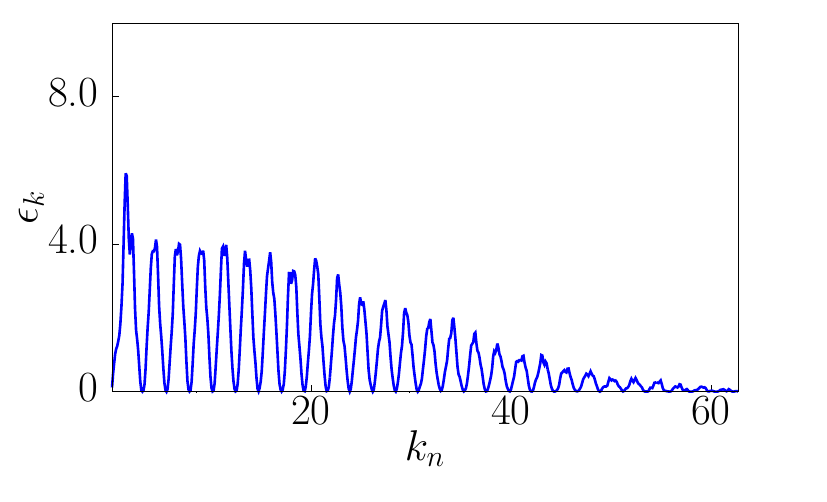}}
\put(220,140){\includegraphics[angle=0,width=0.47\textwidth]{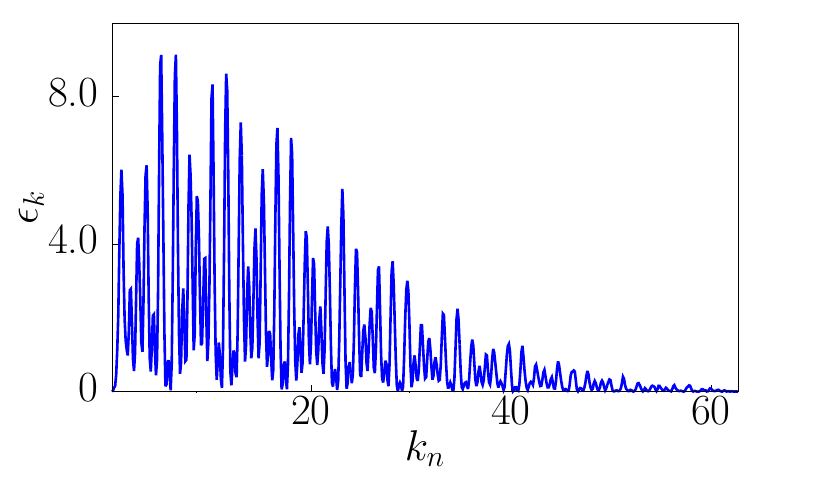}}
\put(20,140){\includegraphics[angle=0,width=0.47\textwidth]{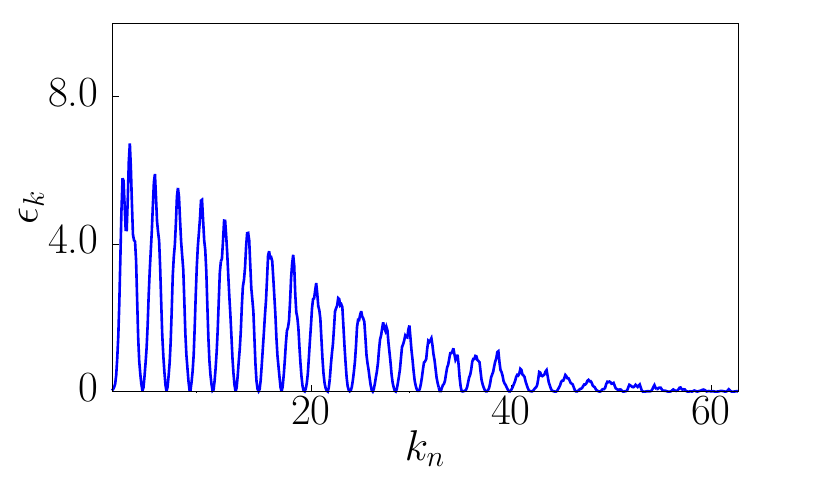}}
\put(220,260){\includegraphics[angle=0,width=0.47\textwidth]{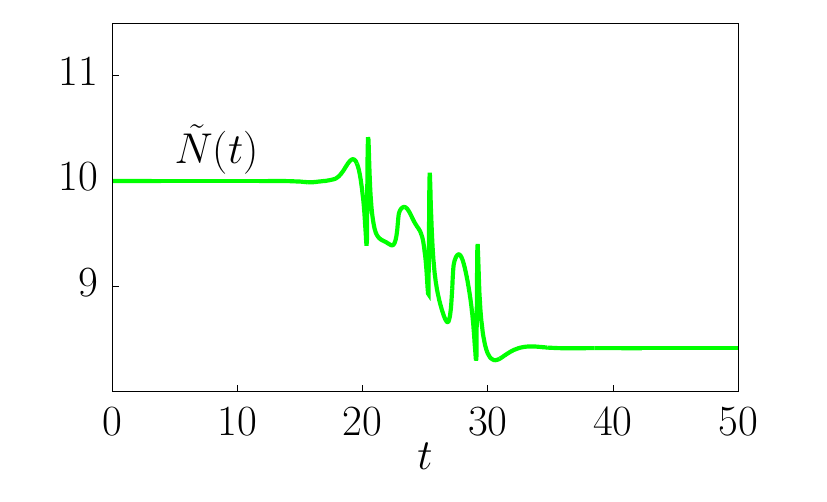}}
\put(20,260){\includegraphics[angle=0,width=0.47\textwidth]{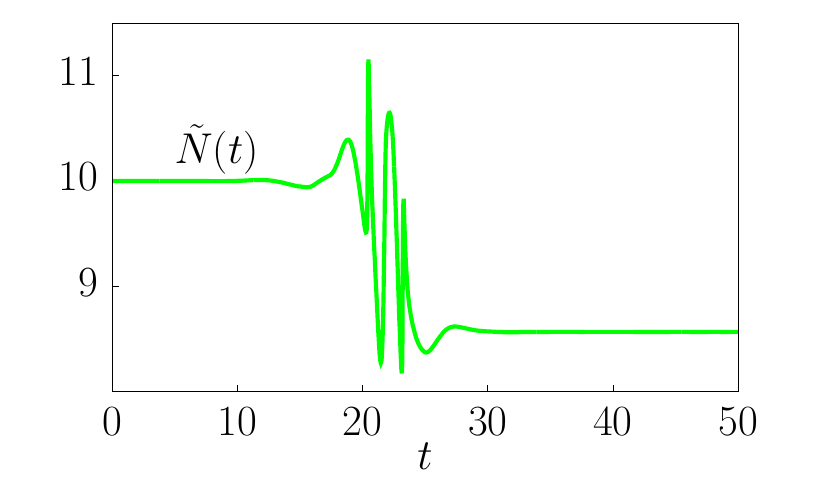}}
\put(220,380){\includegraphics[angle=0,width=0.47\textwidth]{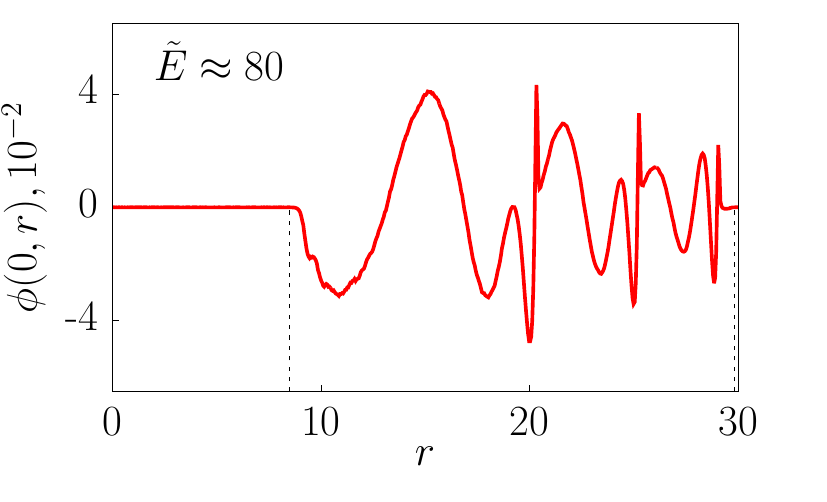}}
\put(20,380){\includegraphics[angle=0,width=0.47\textwidth]{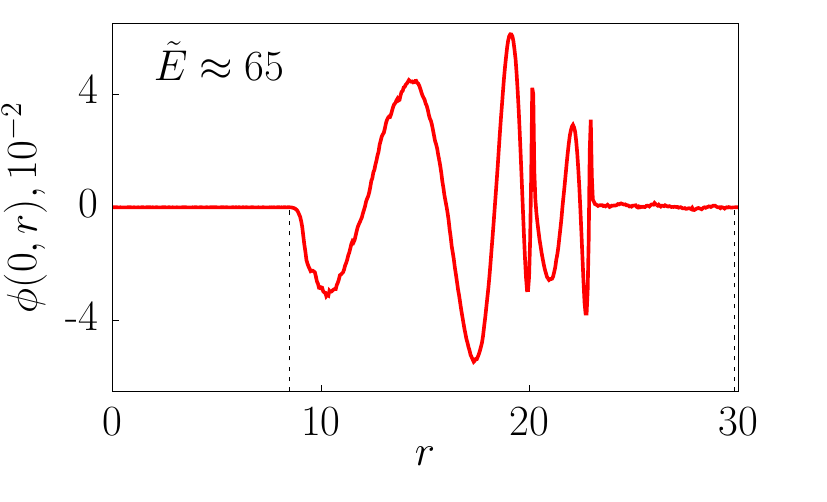}}
\end{picture}
\caption{\label{fig_examples_N10} From top to bottom: a) initial wave
  packet; b) evolution of the instantaneous particle number
  $\tilde{N}(t)$; c) distribution of energy over the modes $\epsilon_k$
  per wave number for the initial field configuration; d) the same as in
  c) but for the final field configuration. Left and right panels correspond to the
  lower boundary solutions with $\tilde{E}\approx65$ and 80, respectively.}
\end{figure}

Let us now turn to the case $\tilde{N}_i=10$ in which we obtain
numerical results concerning boundary
solutions which are qualitatively different from those obtained at
smaller values of $\tilde{N}_i=0.1, 1.0$. In Fig.~\ref{fig_N10}
boundary of the classically allowed region is shown with the thick
solid (red) and thick dashed (blue) lines for the energy interval
$\left[1.5\tilde{N}_i,15.0\tilde{N}_i\right]$ with the same
set~\eqref{4_1} of $\tilde{E}_*$. We find that although 
the form of the boundary is similar to that of for
$\tilde{N}_i=0.1, 1.0$, its upper and lower parts, i.e.
$\tilde{N}_f^{max}(E)$ and $\tilde{N}_f^{min}(E)$, consist of two
different branches of the classical solutions. At energies lower than
about 70 the boundary solutions have two distinct spiky parts. An
example of the initial field configuration $\phi(t=0,r)$ corresponding to
the lower boundary is presented in Fig.~\ref{fig_examples_N10} 
(upper left panel) for $\tilde{E}\approx 65$. The evolution of the instantaneous
particle number $\tilde{N}(t)$ as well as the distribution of energy over
the modes $\epsilon_k$ for initial and final field configurations are
also shown in this figure on left panels. The oscillating pattern in
$\epsilon_k$ indicates that the incoming (and outgoing) field
configuration looks as a sum of two separated in space wavetrains each
having a similar smooth Fourier image. In particular, the distance
$\delta r$ between the spikes in the initial configuration  and the
oscillation period $\delta k$ in the initial energy distribution are
related approximately by $\delta r\cdot \delta k\approx 2\pi$ for all
solutions of 
this branch. For example, the solution with $\tilde{E}\approx 65$
presented on left panels of Fig.~\ref{fig_examples_N10} has
$\delta r\approx 2.8$ and $\delta k\approx 2.3$. Comparing the energy
distributions $\epsilon_k$ for the field configurations at initial and
final times one observes a small energy transfer from low to high
frequency modes. At energies larger
than about 70 our numerical results show that the absolute minimum
(and maximum)  of $\tilde{N}_f$ is delivered by a different branch of
the solutions whose initial (and final) space field configuration contains
already three spikes. On the right panels in
Fig.~\ref{fig_examples_N10} we present the initial field configuration
$\phi(t=0, r)$, time evolution of $\tilde{N}(t)$, and energy
distributions over the Fourier modes $\epsilon_k$ for the initial and
final field configurations for a boundary solution with
$\tilde{E}\approx80$ 
having three spikes. On this branch of solutions the initial and final
field configurations looks as a sum of three separated in space
wavetrains with similar Fourier image. As in the cases of smaller
$\tilde{N}_i$, the instantaneous particle number $\tilde{N}(t)$
undergoes the most dramatic changes when spiky parts of the field
configurations reach the interaction region.
We find that the two-spike
solutions at $\tilde{E}\gsim 70$ and the three-spike solutions at
$\tilde{E}\lsim 70$ still represent local minima of
$\tilde{N}_f$. To figure that out  we take a two-spike lower boundary 
solution at $\tilde{E}\approx65$ as a seed for our Monte-Carlo
procedure 
with $E_*> 70$. To ensure that the field configuration does not
jump to another (three-spike) branch we take a relatively small value of
$a$ which governs the size of changes in the amplitudes $f_n$.
In this way we find that two-spike branch of solutions at
$\tilde{E}\gsim 70$ which is shown in Fig.~\ref{fig_N10} by thin red
(solid) line.  In similar way we find a continuation of the three-spike
branch of the solutions at $\tilde{E}\lsim 70$ which is shown by this
blue (dashed) line in Fig.~\ref{fig_N10}. Upper boundary solutions
have very similar properties\footnote{As previously, for
construction of the boundaries $\tilde{N}^{min}_f(E)$ and
$\tilde{N}^{max}_f(E)$ we select solutions with the longest
oscillating tale, although we obtained also time-shifted field
configurations with the same number of the spiky parts.}. Let us note that in the
chosen energy range the maximal changes in particle number reach
values around 20\%.

\begin{figure}[!htb]
\begin{picture}(300,240)(0,20)
\put(220,20){\includegraphics[angle=0,width=0.47\textwidth]{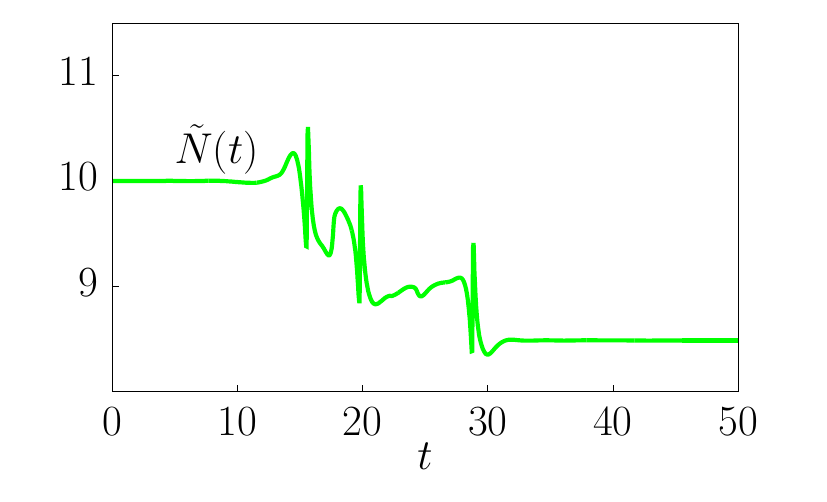}}
\put(20,20){\includegraphics[angle=0,width=0.47\textwidth]{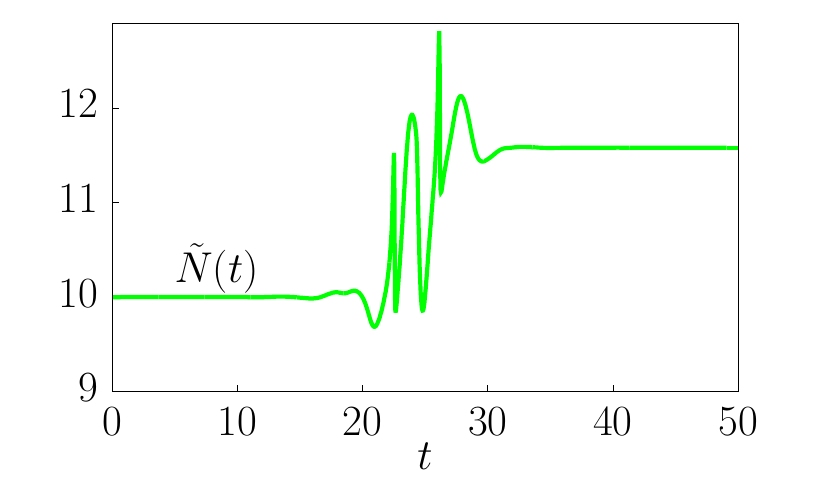}}
\put(220,140){\includegraphics[angle=0,width=0.47\textwidth]{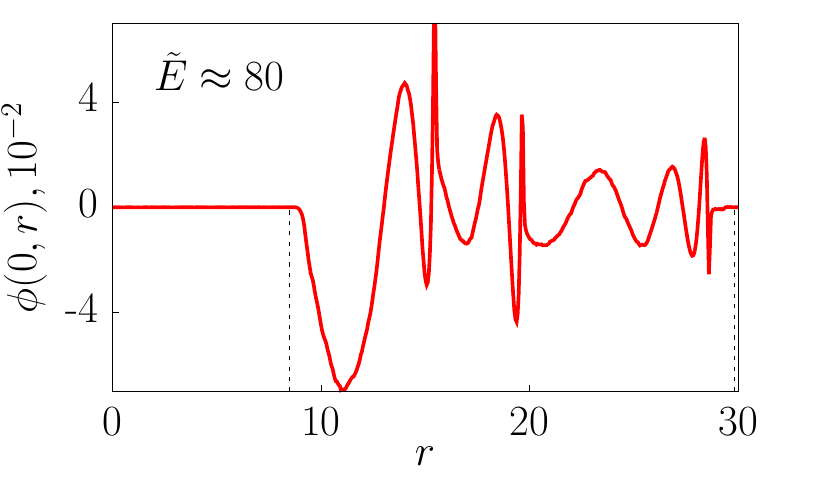}}
\put(20,140){\includegraphics[angle=0,width=0.47\textwidth]{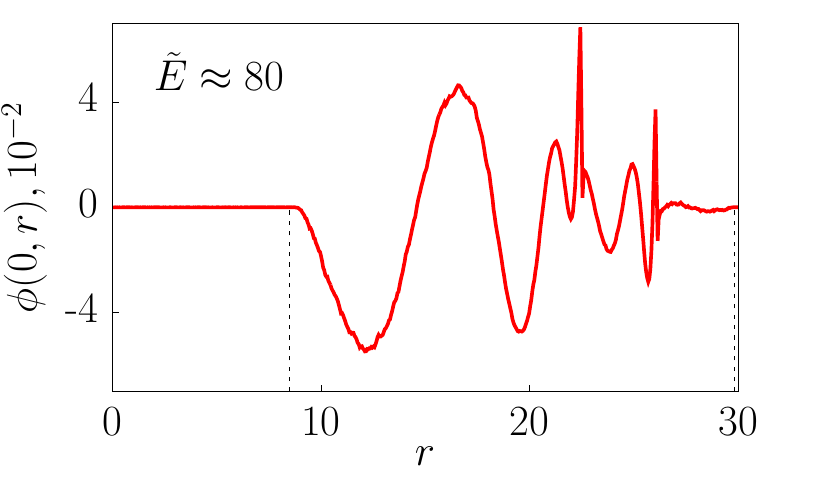}}
\end{picture}
\caption{\label{fig_examples_N10_noboundary} The initial wave packets
  (left panels) and evolution of the instantaneous particle number
  (right panels) for the solutions with $\tilde{N}_i=10$ and
  $\tilde{E}\approx 80$ corresponding to local minima of
  $\tilde{N}_f$. Upper  and lower panels correspond to $\tilde{N}_f>\tilde{N}_i$ and
  $\tilde{N}_f<\tilde{N}_i$. } 
\end{figure}

There have been also found another branches of solutions
which deliver only local extrema to $\tilde{N}_f$. In the linear regime
these solutions look as a sum of two or three single-spike wavetrains 
which differ from those already described by somewhat different distances
between the wavetrains. In Fig.~\ref{fig_examples_N10_noboundary} we show
two examples of such solutions.
Left panels show a solution of this type near the upper boundary with two spikes and
$\tilde{N}_f>\tilde{N}_i$. The distance between the wavetrains in the
initial field configuration is larger than that of for the boundary
two-spike solution. Similarly, the right
panels show an example of the solution near the lower boundary delivering a local minima to
$\tilde{N}_f$ with three spike. Again we observe somewhat larger distance
between the wavetrains in the initial field configuration,
c.f. Fig.~\ref{fig_examples_N10} (upper right panel). These two 
branches of solutions are shown in Fig.~\ref{fig_N10} by dots.
Let us note, that for $\tilde{N}_i=10$ we do not find
any field configuration which would deliver (even local) extrema to
the functional $\tilde{N}_f(E)$ and whose initial (final) wave packet
would have a single spike only. We tried to find it on purpose by taking as
a seed some initial field configuration which was obtained for
$\tilde{N}_i=1$ as a boundary solution which is of single-spike
type and increasing $\tilde{N}_i$ by small steps to
$\tilde{N}_i=10$. We find that at intermediate values of $\tilde{N}_i$
around 5--6 the initial field configurations of the boundary
configurations prefer to be splitted into more than one
wavetrains to produce larger change in particle number
$\left|\tilde{N}_f - \tilde{N}_i\right|$.

\begin{figure}[!htb]
  \begin{center}
    \includegraphics[height=8.0cm]{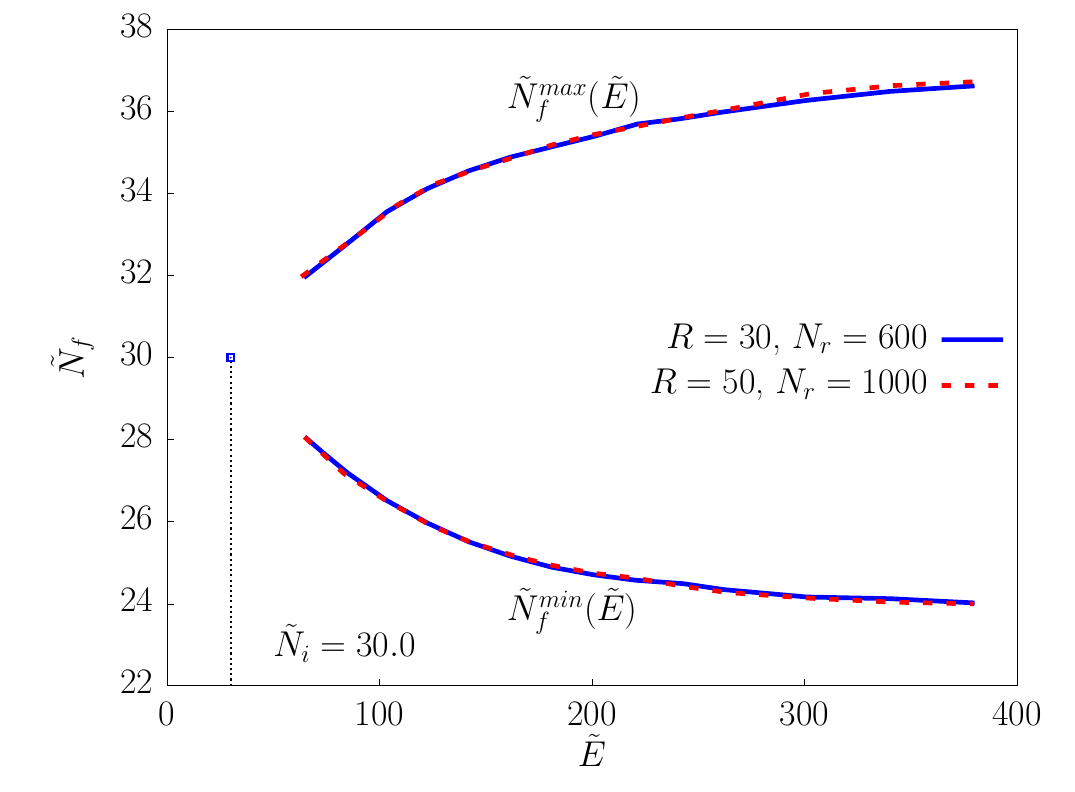}
  \end{center}
\caption{The classically allowed region in the $(\tilde{E}, \tilde{N}_f)$
  plane for $\tilde{N}_i=30.0$. The lattice parameters are $R=30,
  N_r=600$ (solid blue line) and $R=50, N_r=1000$ (dashed red
  line).\label{fig_N30}}  
\end{figure}

\begin{figure}[!htb]
\begin{picture}(300,240)(0,20)
\put(220,20){\includegraphics[angle=0,width=0.47\textwidth]{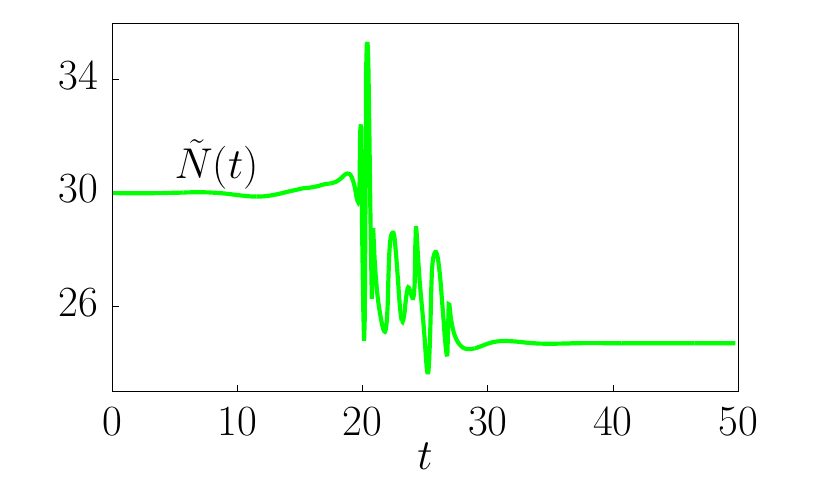}}
\put(20,20){\includegraphics[angle=0,width=0.47\textwidth]{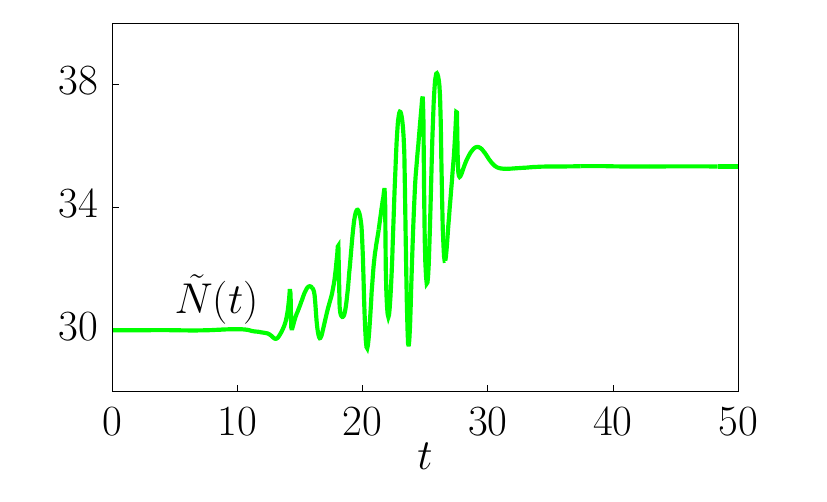}}
\put(220,140){\includegraphics[angle=0,width=0.47\textwidth]{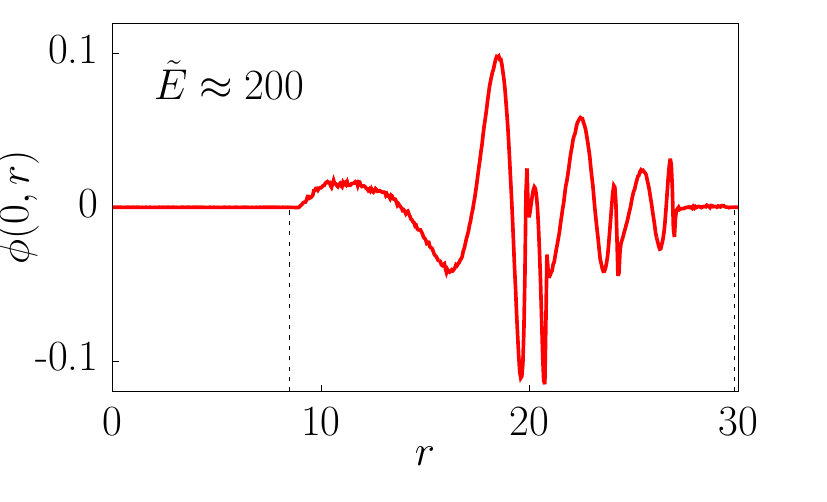}}
\put(20,140){\includegraphics[angle=0,width=0.47\textwidth]{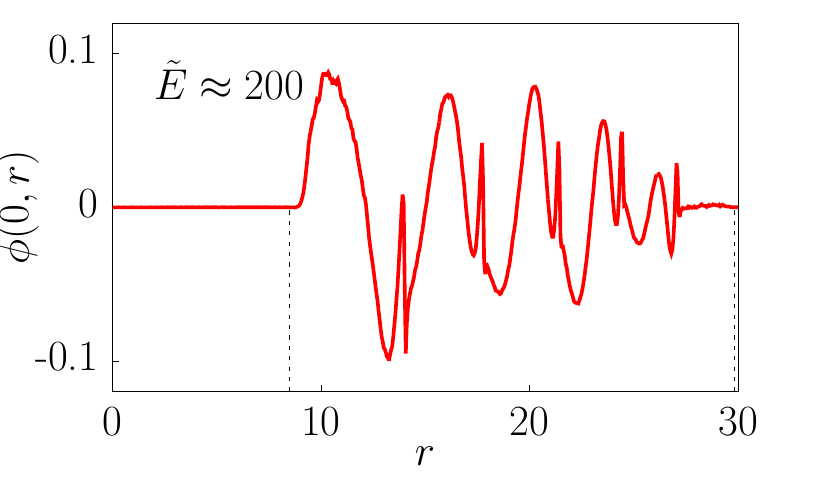}}
\end{picture}
\caption{\label{fig_examples_N30_boundary} The initial wave packets (left panels) and
  evolution of the instantaneous particle number (right panels) for
  the boundary solutions with $\tilde{N}_i=30$ and
  $\tilde{E}\approx 200$. Upper
  and lower panels correspond to $\tilde{N}_f>\tilde{N}_i$ and
  $\tilde{N}_f<\tilde{N}_i$. } 
\end{figure}

The picture becomes even more complicated at larger $\tilde{N}_i$. In
Fig.~\ref{fig_N30} 
we show our numerical results for the boundary of the classically allowed
region  for $\tilde{N}_i=30$ taking $R=30, N_r=600$ (solid blue
line). We observe that the difference $|\tilde{N}_f-\tilde{N}_i|$ does
not exceed 22\% in this case in the chosen energy interval. We find 
that the corresponding boundary solutions contain already 4--7 spikes 
in their initial (and final) wavetrains. Examples of the initial field 
configurations of the boundary solutions are shown in
Fig.~\ref{fig_examples_N30_boundary}. 
At the same time we find that the number of local extrema of $\tilde{N}_f$
increases dramatically at large $\tilde{N}_i$ . Corresponding solutions
have similar number of wavetrains but with different distances
between them. This greatly complicates the task of finding the
boundary of the classically allowed region because, on the one hand,
many of our numerical runs get stuck in such local minima and, on the
other, the values of $\tilde{N}_f$ at the same $\tilde{E}$ for
different branches of near-boundary solutions are at some cases seem
to be close to our numerical accuracy. For these reasons at present we
are unable to classify in details the boundary solutions as it has
been done for the cases with smaller $\tilde{N}_i$. In the
Fig.~\ref{fig_N30} we present only the envelopes
$\tilde{N}_f^{min}(\tilde{E})$ and $\tilde{N}_f^{max}(\tilde{E})$ of
the classically allowed region. 
We performed an additional check by finding $\tilde{N}_f^{max}$ and
$\tilde{N}_f^{min}$ for larger space interval $R=50$ and taking
$N_r=1000$. Corresponding boundary is shown in
Fig.~\ref{fig_examples_N30_boundary} by dashed red line. From our 
numerical results we expect that the number of closely separated spiky
parts in the wavepackets of the boundary solutions continue to grow with 
the increase of $\tilde{N}_i$. 

\begin{figure}[!htb]
  \begin{center}
    \begin{tabular}{cc}
      \includegraphics[height=5.5cm]{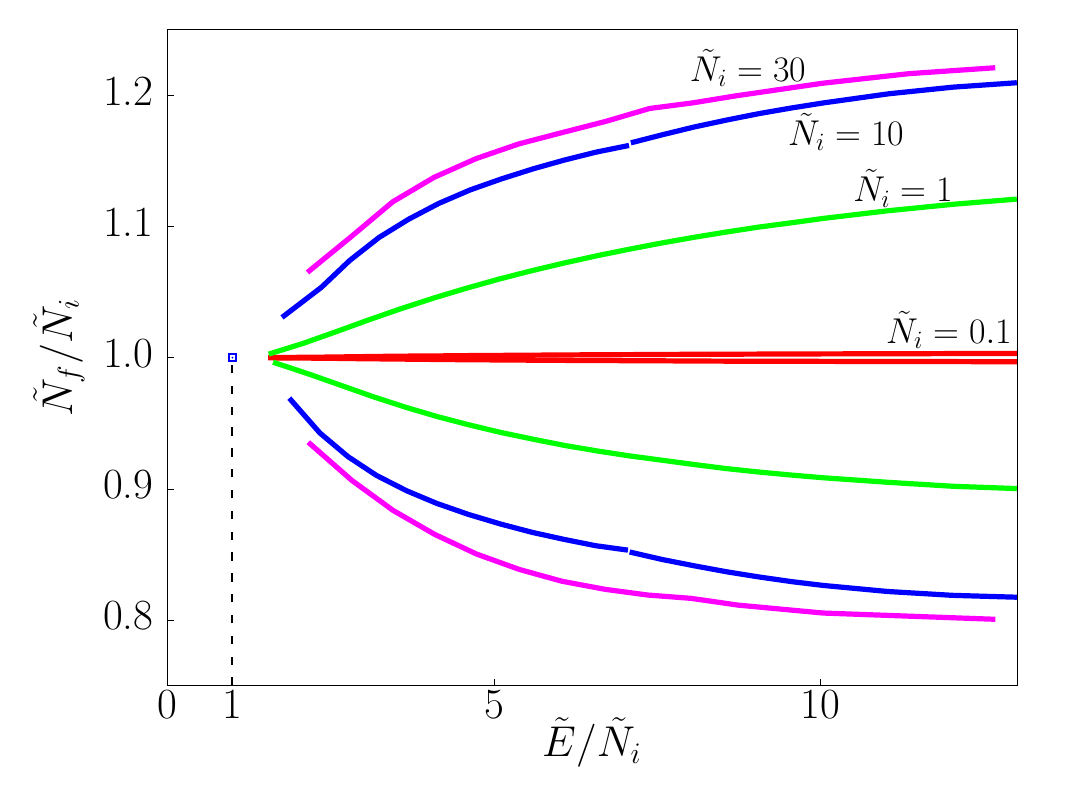} &
      \includegraphics[height=5.5cm]{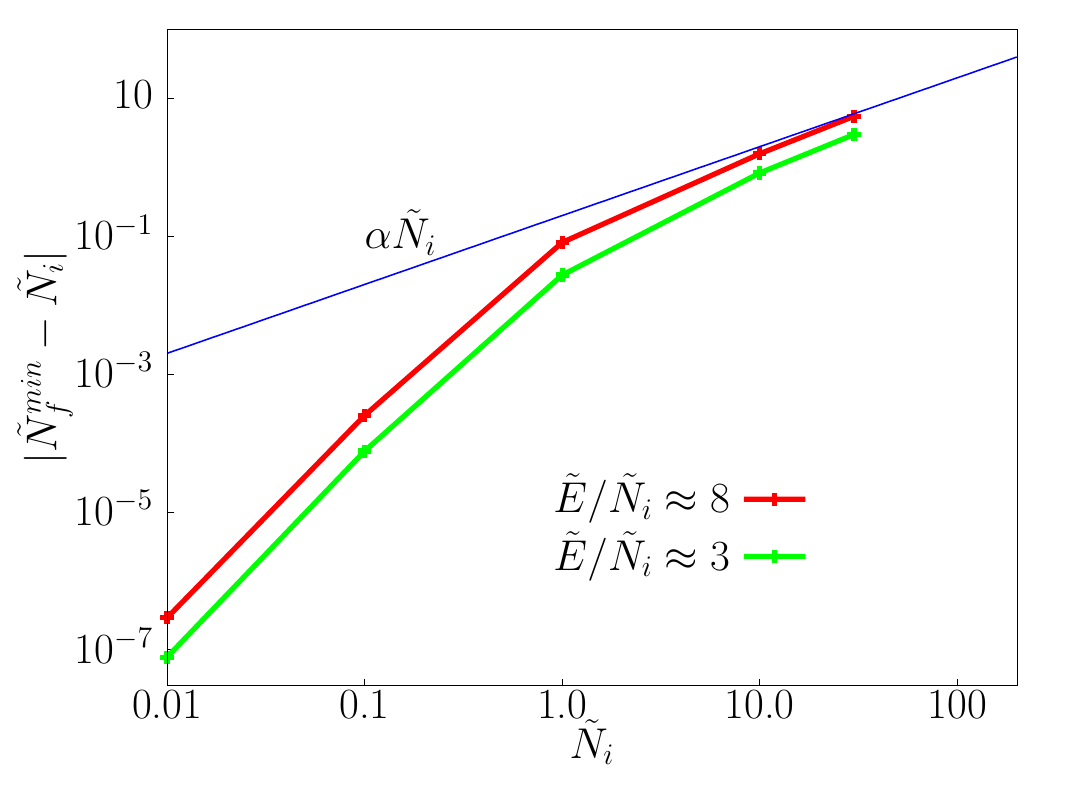}
    \end{tabular}
  \end{center}
\caption{  Left panel: the classically allowed regions in $(\tilde{E},
  \tilde{N}_f)$ plane for different $\tilde{N}_i$. Right panel:
  the difference $|\tilde{N}^{min}_f-\tilde{N}_i|$ as a 
function of $\tilde{N}_i$. \label{fig_scaling}} 
\end{figure}

It is interesting to compare the classically allowed regions at
different $\tilde{N}_i$, see Fig.~\ref{fig_scaling} (left
  panel). Firstly, let us see that if two classical 
solutions describing scattering of waves with different sets of
parameters, i.e. $\tilde{N}^{(1)}_i, \tilde{N}^{(1)}_f,
\tilde{E}^{(1)}$ and $\tilde{N}^{(2)}_i, \tilde{N}^{(2)}_f, 
\tilde{E}^{(2)}$, exists then there should also exist a 
solution which has energy and particle numbers equal to the following
sums 
\be
N_i = N_i^{(1)} + N_i^{(2)}, \;\;
N_f = N_f^{(1)} + N_f^{(2)}, \;\;
E = E^{(1)} + E^{(2)}\,.
\ee
Such a solution can be constructed explicitly by taking it as a
sum of the individual solutions sufficiently separated in
space-time. This observation, in particular, means that the width of
the classically allowed 
region, i.e. $|\tilde{N}^{max}_f-\tilde{N}_i|$ or
$|\tilde{N}^{min}_f-\tilde{N}_i|$ at fixed ratio $\tilde{E}/\tilde{N}_i$
should grow faster than a linear function with increase of initial
particle number $\tilde{N}_i$. On the Fig.~\ref{fig_scaling} we
show\footnote{To plot this dependence we additionally found several
  extrema at the lower boundary for $\tilde{N}_i=0.01$.}
the dependence of the quantity $|\tilde{N}^{min}_f-\tilde{N}_i|$ as a
function of $\tilde{N}_i$ for two values of $\tilde{E}/\tilde{N}_i$. 
For comparison a linear function is shown in thin (blue) line. One can
see that the energy dependence of $|\tilde{N}^{min}_f-\tilde{N}_i|$
tends to be linear at large $\tilde{N}_i$. We found that
$|\tilde{N}^{max}_f-\tilde{N}_i|$ has similar behaviour. This can be
considered as an indication on the existence of some limiting boundary of
the classically allowed region in the plane $(\tilde{E}/\tilde{N}_i,
\tilde{N}_f/\tilde{N}_i)$ at large $\tilde{N}_i$ or in the plane
$(\tilde{E}/\tilde{N}_f, 
\tilde{N}_i/\tilde{N}_f)$ at large $\tilde{N}_f$  due to the symmetry
$\tilde{N}_i\leftrightarrow \tilde{N}_f$.

\section{Discussions and conclusions}
In this paper we considered the classical scattering of wavepackets in the
unbroken $\lambda\phi^4$ theory. Initial and final field
configurations were characterized by energy $E$, initial $N_i$ and 
final $N_f$ particle numbers. We found that although the maximal change of
particle number in the scattering grows with energy at fixed value of
$N_i$ (or $N_f$), the value of $\frac{|N_f-N_i|}{N_{i}}$ does not
exceed 22\% for the considered energy range $E \lsim 10 N_fm$.
In the quantum counterpart of the problem the initial and final wavepackets
correspond to initial and final coherent states with given average
values of energy and particle number. Our classical approach to the
multiparticle scattering is valid when occupation numbers are large,
i.e.  $E, N_i, N_f\sim \frac{1}{\lambda}$ and $\lambda\ll 1$. In this
study we numerically obtained the classically allowed regions of these
processes in the $(E,N_f)$ plane at several values of $N_i$ (or,
equivalently, in the $(E,N_i)$ plane at several values of $N_f$). They
are shown in
Figs.~\ref{fig_N1_Nx400_all},~\ref{fig_N0_1},~\ref{fig_N10}
and~\ref{fig_N30}. Here we considered spherically symmetric field
configurations only. We believe that this simplified consideration has
much in common with most general setup. 

The scattering processes with the parameters $E, N_i$ and $N_f$
outside the classically allowed regions are classically forbidden. The
common lore here is that their probability in quantum theory is
expected to be exponentially suppressed, i.e. have the form ${\cal 
  A}e^{\frac{1}{\lambda}F}$ with some negative suppression exponent
$F$ and a prefactor ${\cal A}$. At the same time the probability of 
the classically allowed processes is not exponentially suppressed. In
the most interesting case $2\to N_f$ production, with $\lambda
N_f\gsim 1$, the initial state contains semiclassically small number
of particles. Our results show that such processes (as well as $N_f\to
2$ scattering) lie deeply in the classically forbidden region and
therefore their probabilities are expected to be exponentially
suppressed at least for $\frac{4\pi}{\lambda}N_f\lsim 30$ and $E\lsim
10 N_f$. Limitations of our numerical procedure do not permit us to
obtain the boundary of the classically allowed region for arbitrarily
large values of $N_f$ and $E$. However, we obtain an evidence for
existence of a limiting boundary of the classically allowed region in
$(E/N_i,N_f/N_i)$ plane at large $N_i$.

 In summary, our results imply that the probability of the
  multiparticle production processes $few\to N_f$ in the unbroken weakly
  coupled $\lambda\phi^4$ theory is suppressed even at large $\lambda
  N_f\gsim 1$. Therefore, factorial growth of multiparticle amplitudes
  observed at $\lambda N_f\ll 1$ should stop at its larger values
  leaving the probability consistent with unitarity. Based on 
  the perturbative results at small $\lambda N_{f}$ and on 
    the experience
  with other non-perturbative processes such as instanton-like
  transitions, false vacuum decay, soliton pair production in particle
  collisions we expect this suppression most likely to be exponential in
  the sense of Eq.~\eqref{0_1}. Let us note that the analysis
  described in this study can be (at a cost of a considerably larger CPU
  time) extended to the case of field configurations with axial
  symmetry which would be more appropriate for discussion of the limit
  of two-particle collisions. Also, it would be interesting to apply the
  method of classical solutions to the scalar theory with the
  spontaneously broken symmetry which we leave for future study. 

Although our findings indicate that $few \to N_f$ scattering processes
lie in the classically forbidden region, they tell nothing about
actual value of the probability of these processes. As we already
mentioned in the introduction there exist semiclassical methods to
calculate the suppression exponent of the probability which, however,
are difficult to apply due to singular nature of corresponding
classical solutions. On the one hand, one can start, as we did in this
paper, with processes $N_i\to N_f$ where both initial and final states
contain large occupation numbers and apply the same semiclassical
techniques. In this case based on the results of the semiclassical
studies of baryon number violating
processes~\cite{Bezrukov:2003er,Bezrukov:2003qm} and soliton pair
production~\cite{Levkov:2004tf,Demidov:2011dk} in particle collisions
one can expect that corresponding classical solutions will be
nonsingular. The classical solutions at the boundary of the classically
allowed region obtained in the present study are expected to be 
close to the solutions describing transitions in the forbidden
region. Using the Rubakov-Son-Tinyakov
conjecture~\cite{Rubakov:1992ec} the probability of $2\to N_f$
scattering in the leading semiclassical approximation can be obtained
from that of $N_i\to N_f$ process by taking the limit $\lambda N_i\to
0$ if it exists\footnote{This conjecture was checked in several
models~\cite{Tinyakov:1991fn,Mueller:1992sc,Bonini:1999kj,Levkov:2008csa}
and was used in
Ref.~\cite{Levkov:2004tf,Demidov:2011dk,Demidov:2015bua,Demidov:2015nea}
for semiclassical calculations of the probability of soliton pair
production and false vacuum decay induced by particle
collisions.}. This procedure can be viewed as a regularization to the
semiclassical method of singular classical solutions. 
We are going to pursue this idea in future work~\cite{future}.

\paragraph*{Acknowledgments}
We thank Fedor Bezrukov, Dmitry Levkov, Alexander Panin, Valery
Rubakov and Sergey Sibiryakov for helpful discussions. The work was
supported by the RSCF grant 14-22-00161. The numerical part of the
work was performed on Calculational Cluster of the Theory Division of
INR RAS. 

\bibliographystyle{JHEP}
\bibliography{classical}

\end{document}